\documentclass[twocolumn]{aastex631}

\usepackage{float}
    
\begin{document}

\title{Different Inhomogeneous Evolutionary Histories for Uranus and Neptune}

\correspondingauthor{Roberto Tejada Arevalo}
\email{arevalo@princeton.edu}

\author[0000-0001-6708-3427]{Roberto Tejada Arevalo}
\affiliation{Department of Astrophysical Sciences, Princeton University, 4 Ivy Lane,
Princeton, NJ 08544, USA}

\begin{abstract}

We present updated non-adiabatic and inhomogeneous evolution models for Uranus and Neptune, employing an interior composition of methane, ammonia, water, and rocks. Following formation trends of the gas giants, Uranus and Neptune formation models are applied, where both planets begin with layers stable to convection. Both planets are subject to convective mixing throughout their evolution. Consistent with past work on this subject, the interior heat of Uranus evolution models is preserved by the stability of an outer composition gradient at lower initial entropy, where convective mixing is inhibited over evolutionary timescales. In contrast, if Neptune's initial entropy is enough to convectively mix its envelope, it undergoes homogenization and adiabatic cooling of the outer 40\% of its envelope. The subsequent release of internal energy during Neptune's evolution, driven by the convective instability of its primordial outer compositional gradient, accounts for its higher luminosity relative to Uranus. This work proposes that the observed luminosity differences between Uranus and Neptune could be explained by the convective stability of their outer envelopes. The extensive convective mixing in Neptune can lead to a higher metallicity in its outer region compared to Uranus, a feature seen in atmospheric measurements and shown in past interior models of Neptune. Due to Neptune's more pronounced cooling, our models predict favorable conditions for hydrogen-water immiscibility in its envelope. 

\end{abstract}


\section{Introduction}

Uranus and Neptune remain the least understood gas giant planets in our Solar System. Unlike Jupiter and Saturn, extensively studied by spacecraft such as \textit{Juno} \citep{Bolton2017a} and \textit{Cassini} \citep{Matson2003}, Uranus and Neptune have been visited only once by \textit{Voyager 2}, in 1986 and 1989, respectively. Despite the limited observational data, valuable insights have been gained into the internal structures of Uranus and Neptune. Over the past three decades, exoplanet research has uncovered that most of the exoplanet masses and sizes in our galaxy are between the size of the Earth and Neptune \citep{Winn2015, Dressing2015a, Christiansen2023, Parc2024}, with recent exoplanet population studies finding that Neptune-sized objects are among the most common \citep{Fulton2018a} and can serve as calibrations for exoplanet observations \citep{Hasler2024}. Hence, understanding their formation, interior structure, and evolution is becoming ever more relevant.

The difference between the luminosities of Uranus and Neptune is one of the most significant questions regarding both planets. Early studies indicated that Neptune's luminosity could be accounted for using relatively simple adiabatic models \citep{Hubbard1978}. At the same time, similar early three-layer models for Uranus predicted cooling times longer than those implied by its measured effective temperature \citep{Hubbard1978, Hubbard1995, Fortney2011, Scheibe2019}. Consequently, the near-zero internal flux observed in Uranus remained a longstanding puzzle \citep{Pearl1991, Podolak1995, Hubbard1995, Fortney2011}, particularly given the similar mass, radius, atmospheric metallicity, and magnetic field characteristics of the two planets \citep{Guillot2015}. Traditionally, evolutionary models for Uranus and Neptune have employed a three-layer adiabatic structure consisting of a rocky core, a water-rich envelope, and an outer hydrogen-helium (H-He) layer \citep[e.g.,][]{Hubbard1968, Hubbard1977, Hubbard1978, HubbardMacFarlane1980, Nettelmann2013a}. To address Uranus’s low luminosity, \cite{Nettelmann2016} introduced a thermal boundary layer (TBL), and positioned it in the outer layers of the planet. This TBL restricts internal heat transport, thereby enabling efficient cooling of the outer envelope while conserving heat within the deep interior \citep[also see][]{Scheibe2019, Scheibe2021}.

The gravitational harmonic measurements of Uranus by \textit{Voayger 2} indicated incompatibility with strictly adiabatic interiors \citep{Podolak1995}, prompting \citet[][VH20]{VazanHelled2020} to investigate inhomogeneous, non-adiabatic evolution models of Uranus with gradual primordial composition gradients. Their findings demonstrated that sufficiently steep outer heavy element composition ($Z$) gradients stabilized the outer envelope against convection \citep{Ledoux1947}, effectively maintaining low luminosity over evolutionary timescales. The Uranus models of VH20, characterized by limited convective mixing and steep outer gradients, supported a model with a mixture of water and rock at high temperatures without distinct layering and hydrogen and helium (H-He) confined primarily to the outer envelopes.

Subsequent research suggested that incorporating higher H-He abundances with ammonia-methane-water ice mixtures may further refine interior models of Uranus and Neptune \citep{Movshovitz2022, Lin2024}. Particularly, \cite{Lin2024} found a stronger consistency of ice-H-He mixtures with the gravity data ($J_2,\ J_4$) of Uranus. Independently, formation models by \citet[][VH22]{Valletta2022} also predicted significant mixing of ices with H-He, suggesting H-He fractions between 20\% and 40\% in the planetary envelopes of the nascent Uranus and Neptune.

The work of VH20 emphasized the lack of convective mixing in Uranus's evolution to explain its low luminosity, but left open the question of its impact on Neptune. Recent studies on gas giant evolution \citep[e.g.,][]{Knierim2024, Knierim2025, Tejada2025, Sur2025a} have increasingly recognized the role of convective mixing in shaping primordial composition gradients. These investigations posited enhanced convective mixing at higher interior entropies or with gradients positioned outwardly within the planet. These findings, initially demonstrated by \cite{Knierim2024} using the stellar evolution code \texttt{MESA} \citep{Paxton2011, Paxton2013, Paxton2018}, have been independently validated by \citet{Tejada2025} using the gas giant evolution code \texttt{APPLE} \citep{Sur2024a}.

In this study, we present updated evolutionary models of Uranus and Neptune guided by formation scenarios proposed by VH22. We aim here to explain the luminosity differences between these ice giants via evolutionary processes. Previous hypotheses for the low intrinsic luminosity of Uranus included silicates in Uranus's atmosphere \citep{HubbardMacFarlane1980}, giant impacts \citep{Reinhardt2020}, the presence of a freezing core \citep{Stixrude2021}, water-hydrogen phase separation \citep{Bailey2021}, and enhanced conduction \citep[see discussion in][]{Scheibe2021}. Yet, none have concurrently addressed the history of the radius, effective temperature, luminosity, and gravitational harmonics consistently. While past studies have focused primarily on Uranus, our research here investigates how convective mixing impacts luminosity and effective and internal temperature evolutionary trajectories of both planets, incorporating ammonia-methane-water and rock mixtures throughout their envelopes.

The planetary evolution code, methodology, including equations of state, atmospheric assumptions, and heat and compositional transport processes, are described in Section \ref{sec:methods}. Section \ref{sec:experiments} shows the consequences of different initial thermal profiles on the evolution of a Uranus-mass model with the same heavy element profile, outer metallicity, and core mass. Section \ref{sec:evolution} showcases preferred models of Uranus and Neptune models which explain the luminosity differences while matching the relevant observable parameters. Finally, Section \ref{sec:conclusion} summarizes our key findings and offers concluding remarks.

\begin{deluxetable*}{ccc}
\tablecaption{Uranus and Neptune Parameters\label{tab:params}}
\tablehead{
  \colhead{Parameter} & \colhead{Uranus} & \colhead{Neptune}
}
\startdata
Radius (R$_\oplus$)               & 4.007                   & 3.883                  \\
Mass (M$_\oplus$)                 & 14.536                  & 17.147                 \\
Effective Temperature (K)         & $59.100 \pm 0.3$          & $59.300 \pm 0.8$                \\
$J_2 \times 10^6$                 & $3509.709 \pm 0.064$    & $3408.43 \pm 4.5$      \\
$J_4 \times 10^6$                 & $-35.04 \pm 0.140$     & $-33.40 \pm 2.90$      \\
Rotation Period (V2; hours)       & 17.24                   & 16.11                  \\
Rotation Period (H10; hours)      & 16.579                 & 17.458                  \\
Bond Albedo (PC91)                & $0.300 \pm 0.049 $        & $0.290 \pm 0.067$        \\
Bond Albedo (I25)                 & $0.349 \pm 0.016 $        & -                      \\
Emitted Power (PC91 10$^{16}$ W)   & $0.560 \pm 0.011$       &  $0.534 \pm 0.029$ \\
Absorbed Power (PC91 10$^{16}$ W)   & $0.526 \pm 0.037$       &  $0.204 \pm 0.019$ \\
Internal Power (PC91 10$^{16}$ W)   & $0.034 \pm 0.038$       &  $0.330 \pm 0.035$ \\
Intrinsic Flux (PC91 W m$^{-2}$)   & $0.042 \pm 0.047$       &  $0.433 \pm 0.046$ \\
Intrinsic Flux (W25; W m$^{-2}$)    & $0.078\pm 0.018$       &  - \\
Intrinsic Flux (I25; W m$^{-2}$)    & $0.089\pm 0.030$       &  - \\
\enddata
\tablecomments{Gravity data for Uranus and Neptune come from \cite{Jacobson2025} and \cite{Jacobson2009}, respectively. The effective temperature and Bond albedo measurements are derived by \citet[][PC91]{Pearl1991} from \textit{Voyager 2} data. Likewise, the emitted power, absorbed power, and intrinsic flux was measured by PC91. Recently, \citet[][W25]{Wang2025} measured a higher intrinsic flux from Uranus based on seasonal variations of the planet. A new Bond albedo was found by \citet[I25][]{Irwin2025}, who also found a higher intrinsic flux for Uranus. The rotation periods of both planets are tabulated as measured by \textit{Voyager 2} \citep[V2;][]{Desch1986, Warwick1986, Warwick1989} and as revised by \citet[][H10]{Helled2010}.}
\end{deluxetable*}

\section{Methods}\label{sec:methods}

\subsection{Equations of State}\label{subsec:eos}

We use the hydrogen-helium (H-He) equation of state (EOS) from \cite{Chabrier2021}, which is based on the EOS of \cite{Chabrier2019} and incorporates the non-ideal entropy and density effects of \cite{Militzer2013}. All models in this work have a constant $Y/(X + Y) = 0.277$, consistent with solar abundance ratios \citep{Bahcall2006}, where $X$ and $Y$ are the relative hydrogen and helium mass fractions. In contrast to previous pioneering studies that modeled Uranus and Neptune interiors exclusively with pure water and rocks \citep[][]{Scheibe2019, Scheibe2021, VazanHelled2020}, we adopt a solar composition mixture comprising methane, ammonia, and water in a 4:1:7 ratio \citep{Asplund2009, Bethkenhagen2017} using the volume-addition law \citep[See discussions in][]{Saumon1995, Tejada2024}. \cite{Bethkenhagen2017} found deviations of the volume-addition law up to 4\% from direct mixture EOS calculations, justifying its use here. The impact of different ice ratios is discussed and shown in Appendix A. Recently, evolution models of Uranus and Neptune incorporated H-He and pure water mixtures \citep{Scheibe2019, Scheibe2021} and rocks \citep{VazanHelled2020}. The motivation to use ice mixtures instead of pure water comes from the expectation that ice mixtures are expected in both of these planets since their interior conditions are favorable \citep{Bethkenhagen2017, Cheng2023, Lin2024} and their atmospheric metallicity abundances show evidence for ices \citep{Karkoschka1994, Karkoschka1998}. For the ice mixture, we use the multi-phase water EOS provided by \citet[][;AQUA]{Haldemann2020}, equipped with the updated water EOS corrections from \cite{Mazevet2019}.\footnote{\cite{Aguichine2025} noted that a correction to the entropy values of \cite{Mazevet2019} EOS has been issued by \cite{Mazevet2021}. This correction does not affect the density. We calculated models using a version of AQUA replacing the \cite{Mazevet2019} EOS with the \cite{Mazevet2021} EOS, finding no significant changes to the evolution models and no changes to the broader conclusions presented here. See Appendix A, Figure \ref{fig:fig7}, for a discussion and examples.} The ammonia and methane EOSes are those calculated by \cite{Bethkenhagen2017}. We transition smoothly to the analytic EOS fits computed by \cite{Gao2023} for ammonia and \cite{Setzmann1991} for methane to capture lower temperature and density conditions \citep[below the lower limits of 1000 K and 0.6 g cm$^{-3}$;][]{Bethkenhagen2017}.\footnote{The zero points of the \cite{Gao2023} and \cite{Setzmann1991} EOSes are chosen to be consistent with the zero point of the EOSes of \cite{Bethkenhagen2017}.}

\begin{figure}[ht!]
\centering
\includegraphics[width=0.47\textwidth]{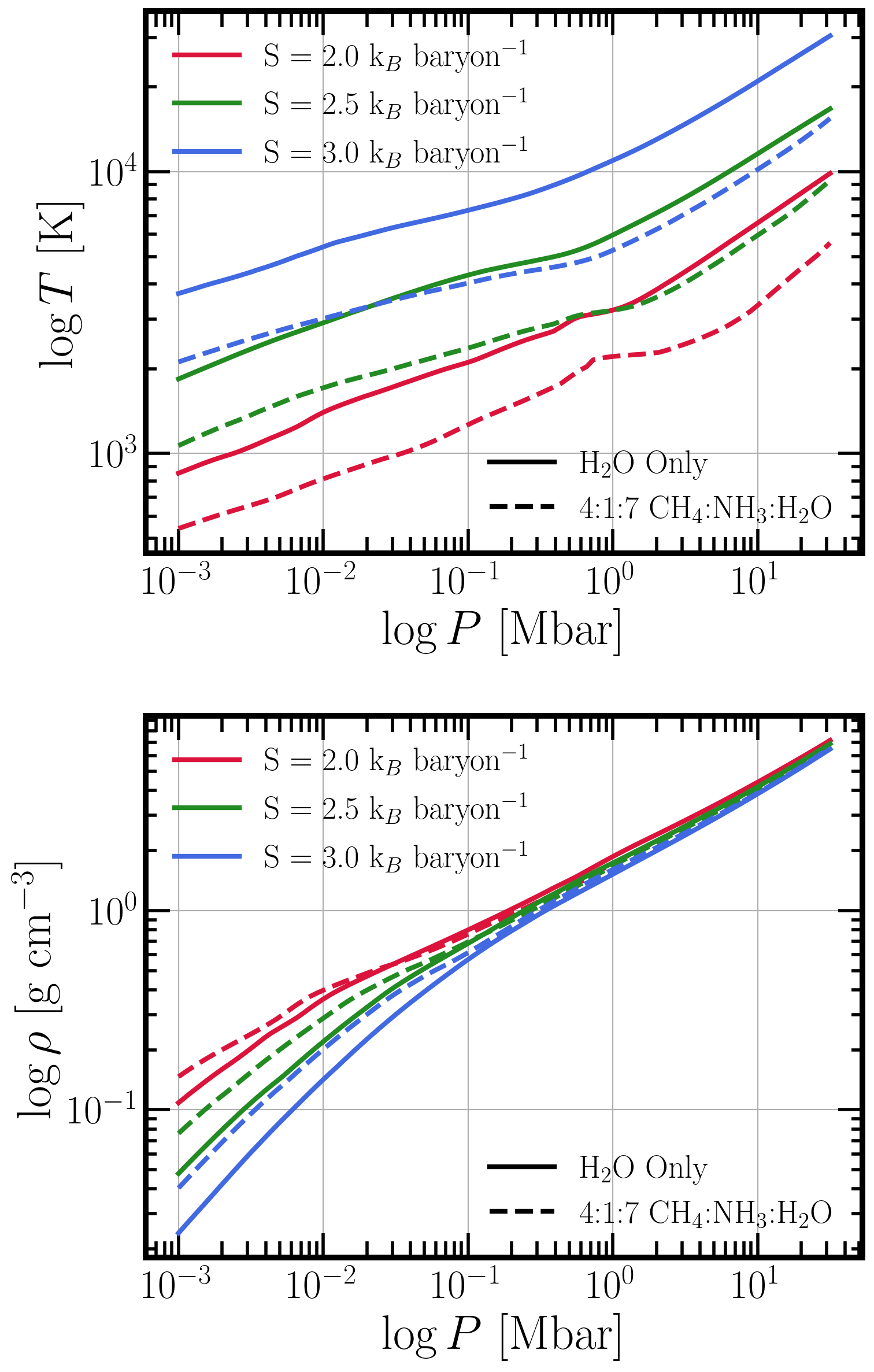}
\caption{Isentropes of EOS mixtures compared at different entropies, denoted by different colors, for $Z = 0.8$ and $Y/(X + Y) = 0.277$ using the \cite{Chabrier2021} H-He EOS. The solid and dashed lines differentiate between H-He water mixtures and H-He-ices mixtures respectively, with the ice mixture being at the measured 4:1:7 solar ratio of methane, water, and ammonia \citep{Asplund2009, Bethkenhagen2017}. The selected pressure range and entropy values are most relevant for the interior of Uranus and Neptune.  Top: The temperature as a function of pressure of a pure water mixture is significantly hotter ($\sim$4,000 K at 10 Mbar) than that of ice mixtures at a particular entropy. Bottom: Densities as a function of pressure. The densities of the ice mixture is greater than the pure water mixture due to their lower temperatures at a constant entropy for pressures lower than $\sim$1 Mbar. These differences show that pure water with H-He is not a close approximation to an ice mixture with H-He.}
\label{fig:fig1}
\end{figure}

Figure \ref{fig:fig1} illustrates the resulting temperature and density differences between H-He-water and H-He-ice mixtures ($Z = 0.8$) at various specific entropies. The isentropic differences in temperature and density shown in Figure \ref{fig:fig1} between H-He-water and H-He-ice mixtures are several $10^3$ K and $\gtrsim 5\%$ in density. These differences are significant enough to warrant the use of ice mixtures instead of pure water. We, hence, argue here that a pure water EOS does not sufficiently approximate ice mixtures to justify using pure water in Uranus and Neptune models. For rocky materials, we use the MgSiO$_3$ EOS originally derived by \cite{Keane1954}, including the vibrational and thermal effects as calculated and applied by \cite{Zhang2022}. We omit considerations of iron content, since current evidence suggests that the ice giants likely possess minimal iron in their cores or envelopes \citep{Movshovitz2022, Malamud2024}. A more detailed discussion of the H-He-Z mixture EOS can be found in \cite{Tejada2024}.

\subsection{Atmospheres}

There are no up-to-date atmosphere models for Uranus and Neptune which agree with the observed temperature at 1 bar \citep{Nettelmann2013a, Nettelmann2016}, the Bond albedo, and the atmospheric metallicity dependence. While we acknowledge that more physically accurate models exist \citep{Fortney2011}, \cite{Nettelmann2013a} showed that the differences between these and the simple analytic fits of \cite{Graboske1975} are not too significant given everything else we do not know about these planets.\footnote{Our own sensitivity tests agree with \cite{Nettelmann2016}.} In the absence of such atmosphere models, we use the analytic atmosphere fits of \cite{Graboske1975}, \cite{Hubbard1977} and \cite{Guillot1995}. 

The effective temperature takes the following form:

\begin{equation}\label{eq:Tint}
    T_{\rm eff} = \bigg(\frac{T_{\rm 1 bar}}{K} g^{1/6}\bigg)^{1/1.244},
\end{equation}

$T_{\rm 1 bar}$ being the temperature at 1 bar and $g = GM/R^2$. Then, the internal temperature is obtained via

\begin{equation}\label{eq:Teff}
    T_{\rm int} = \bigg(T_{\rm eff}^4 - T_{\rm eq}^4\bigg)^{1/4},
\end{equation}
where $T_{\rm eq}$ is the equilibrium temperature,

\begin{equation}\label{eq:Teq}
    T_{\rm eq} = T_\odot(1 - A)^{1/4} \bigg(\frac{R_\odot}{2d}\bigg)^{1/2}
\end{equation}
of which $A$ is the Bond albedo, $d$ is the semi-major axis, and $T_\odot$ and $R_\odot$ are the effective temperature and radius of the Sun respectively. Throughout this work, we apply the coefficients $K_U = 1.481$ and $K_N = 1.451$ in Equation \ref{eq:Tint}, Bond albedos of 0.3 and 0.29 \citep{Pearl1991}, and $d = 19.19$ A.U. and $d = 30.07$ A.U. in Equation \ref{eq:Teq} for Uranus and Neptune, respectively, as done in \cite{Scheibe2021}. The intrinsic luminosity flux of the evolution models is hence calculated via the Stefan-Boltzmann law:

\begin{equation}
    \mathcal{F_{\rm int}} = \sigma_B T_{\rm int}^4,
\end{equation}
where $\sigma_B = 5.6704 \times 10^{-8}$ W m$^{-2}$ K$^{-4}$ is the Stefan-Boltzmann constant. 

\vspace{10mm}

\subsection{Evolution and Transport}

All models computed and presented here are calculated with the gas giant planet evolution code, \texttt{APPLE} \citep{Sur2024a}. Solid-body rotation is applied under a Henyey hydrostatic relaxation scheme \citep{Henyey1964} using the updated rotation rates calculated by \cite{Helled2010}. We calculate the moment of inertia and gravitational harmonics from the model structure using the Theory of Figures to Fourth Order approach \citep[ToF4;][]{Nettelmann2017}. The relative errors between ToF4 and ToF7 are $\lesssim 10^{-5}$ \citep{Nettelmann2021}. This approach is the same as that of \cite{Sur2025a} in the context of Jupiter and Saturn evolution.

The Ledoux criterion for convection \citep{Ledoux1947} is used throughout this work. Convective instability occurs in mass zones where

\begin{equation}\label{eq:ledoux}
    \frac{dS}{dr} - \sum_i\frac{\partial S}{\partial X_i}_{\rho, P}\frac{dX_i}{dr} < 0,
\end{equation}
with $X_i$ representing helium and heavy element mass fractions, and $S$ the total specific entropy. This form of the Ledoux condition is detailed in \cite{Lattimer1981}, \cite{Tejada2024}, and \cite{Knierim2024}. The mixing-length theory model of convection used here is described by \cite{Sur2024a}. A lengthy discussion on our application of the Ledoux condition is found in Section 2 of \cite{Tejada2025}. Semiconvection is ignored in this work due to the already existing uncertainties of the semiconvective fluxes in gas giants generally. Conditions for the onset of semiconvection in ice giants could be unfavorable \citep{French2019a}. 

The conductive heat transport implemented in \texttt{APPLE} is comprehensively described in \cite{Sur2024a}. In this study, we adopt the multi-phase thermal conductivity model of \cite{French2019} for the heavy-element thermal conduction component. Besides thermal conduction, compositional diffusion is calculated via the diffusion equation is detailed in Section 5 of \cite{Sur2024a}.

\cite{Bethkenhagen2017} calculated self-diffusion coefficients for H-C-N-O species along adiabatic profiles, obtaining values ranging from $10^{-5}$ to $10^{-3}$ cm$^2$ s$^{-1}$. However, comprehensive density and temperature-dependent diffusion coefficients for the specific hydrogen, ice, and rock mixtures considered in this work are currently unavailable. Therefore, we employ constant heavy-element self-diffusion coefficients, assigning values of $10^{-5}$ cm$^2$ s$^{-1}$ within the deeper interior regions and $10^{-4}$ cm$^2$ s$^{-1}$ for outer compositional gradient regions. This selection is informed by the diffusion coefficient profiles in \cite{Bethkenhagen2017}.

It should be noted that, since the diffusion coefficient profiles computed along Uranus adiabats in \cite{Bethkenhagen2017}, these are strictly applicable only along those adiabats. Consequently, we adopt uniform diffusion coefficients as described above. Sensitivity tests varying the self-diffusion coefficient of ice mixtures between $10^{-3}$ and $10^{-5}$ cm$^2$ s$^{-1}$ yielded negligible differences in model outcomes. Finally, we use a self-diffusion coefficient for hydrogen-helium mixtures of $10^{-3}$ cm$^2$ s$^{-1}$, an order-of-magnitude estimate consistent with the hydrogen diffusion coefficient shown in Figure 5 of \cite{Bethkenhagen2017}.

\section{Evolution Models With Varying Entropy Profiles}\label{sec:experiments}

\begin{figure}[ht!]
\centering
\includegraphics[width=0.47\textwidth]{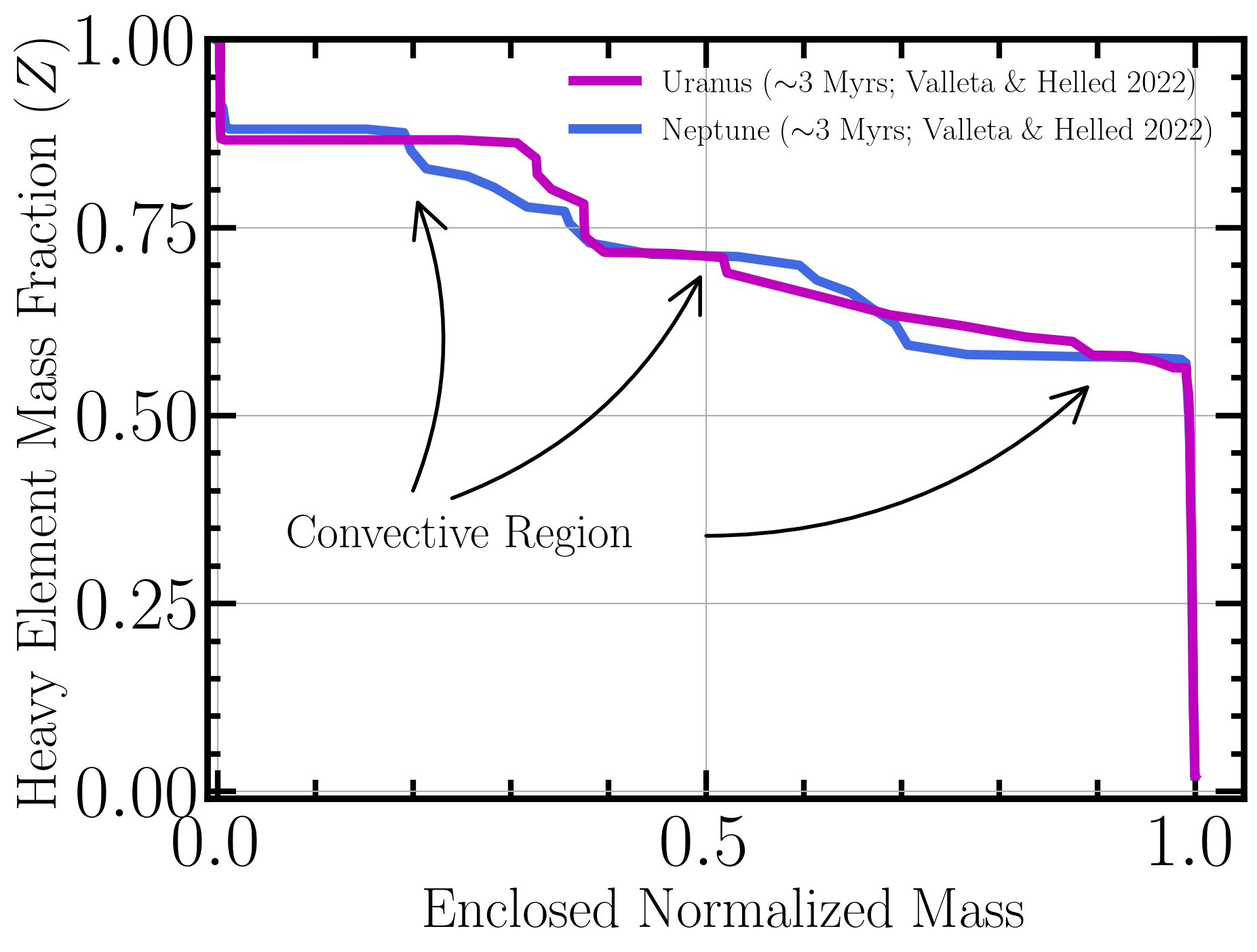}
\caption{Example of post-formation heavy element profiles ($Z$) of Uranus and Neptune from \citet[][VH22]{Valletta2022}, adapted to be shown as a function of the enclosed mass grid from their Figure 1. The formation models of both planets show a steep, outer $Z$ profile gradient. Interior to this, the formation profiles show several stabilizing $Z$ gradients. The homogeneous sections of the Uranus and Neptune formation profiles are likely convective zones (VH22). Guided by these models, we place three convective zones in the initial profiles used in this study. }
\label{fig:fig2}
\end{figure}

This study investigates formation-based initial conditions derived from VH22, examples of which are depicted in Figure \ref{fig:fig2}. The formation models presented there exhibit multiple initial convective layers separated by a pronounced outer compositional gradient, analogous to a TBL. All models in this section have formation-guided initial conditions, characterized by three convective zones and three convectively stable layers as illustrated in Figure \ref{fig:fig2}. Previous studies \citep{Nettelmann2016, VazanHelled2020, Scheibe2021} have examined the evolutionary implications of outer composition gradients and TBLs, generally assuming minimal H-He fractions in the deep interior regions. Here, we relax this assumption, permitting H-He total mass fractions up to $15\%$ in the initial formation-based models. Convective mixing effects are predominantly governed by initial heavy-element and entropy profiles \citep{Knierim2024, Tejada2025}. Stabilizing gradients deeper in the planet's interior show greater stability compared to those positioned further outward, as demonstrated by \citet[][see their Figure 2]{Tejada2025} in the context of Jupiter and generalized by \cite{Knierim2024} for gas giants broadly. Given these established findings, such analyses are not repeated here.

\begin{figure*}[ht!]
\centering
\includegraphics[width=\textwidth]{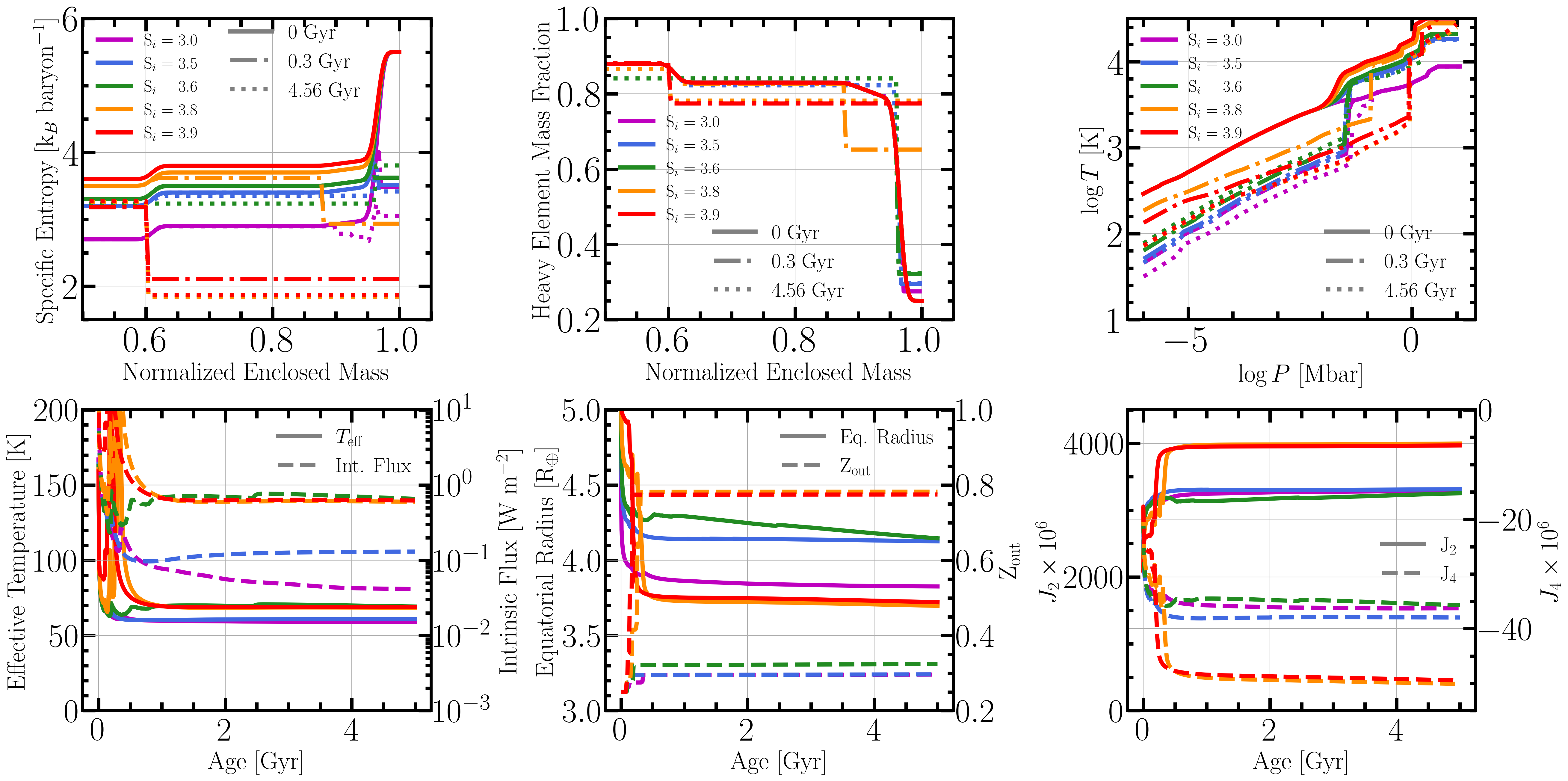}
\caption{Evolution of Uranus-mass models with initial interior entropies ($S_i$) ranging between 3 and 3.9 k$_B$ baryon$^{-1}$. The top row shows the initial and present-age profiles of the entropy (left), heavy element abundance (center), and temperature (right). The bottom row shows the evolution of the effective temperature and intrinsic flux (left), radius, and outer $Z$ (center), and $J_2$, $J_4$ (left). The magenta, blue, and green models preserve an outer $Z$ profile gradient, keeping interior temperatures high, displaying larger radii, leading to low outer intrinsic fluxes. The orange and red model are the hottest and do not preserve an outer composition gradient. The red model mixes almost entirely by 0.3 Gyr and the orange by 0.4 Gyr. This mixing event causes an ephemeral yet dramatic increase in effective temperature, luminosity, and radius. Note that while the green model achieve similarly high final luminosities (center left), it evolves to different radii (green is inflated). This is discussed and further visualized in Appendix B. Models at or higher initial interior entropy than $\sim$3.7 convectively mix the outer regions. The heat escape subsequent to convective mixing yields smaller radii (bottom center) and higher final intrinsic fluxes (bottom right).  The evolution of the outer heavy element abundance of the red model increases from the initial 25\% to 78\% by mass during the mixing event (dashed lines, bottom center). The evolution of $J_2$ and $J_4$ is significantly different from models which do not mix (bottom right).}
\label{fig:fig3}
\end{figure*}

Figure \ref{fig:fig3} illustrates the effects of varying interior specific entropy values from 3 to 3.9 k$_B$ baryon$^{-1}$ in Uranus-mass models, analogous to the approach of Section 3.2 in \cite{Tejada2025} for Jupiter. Recent evolution studies and static interior work have found that Uranus could harbor interior temperatures of tens of thousands of kelvin provided its interior is non-adiabatic \citep{VazanHelled2020, Scheibe2021, Neuenschwander2024, Lin2024}. These initial entropy ranges correspond to a hot Uranian interior, further justifying our choice of interior entropy. All models in this Section are initialized at the same outer entropy of 5.5 k$_B$ baryon$^{-1}$.\footnote{The outermost entropy values have no long-term stabilizing effect, see \cite{Knierim2024} and \cite{Tejada2025} for discussions.} The differences in initial interior entropy profiles lead to distinct evolutionary pathways in temperature and composition, with higher initial entropies promoting extensive convective mixing. The bottom panel of Figure \ref{fig:fig3} displays the radius, effective temperature, luminosity, outer heavy-element abundance, and gravity harmonic evolution of each model. Higher luminosities in these models are attributed to rapid heat release from zones interior to the outer compositional gradient into the external layers, particularly evident in the hottest model (red curve), which undergoes near-total convective mixing. The hottest model undergoes a sudden convective mixing event which causes a dramatic, yet temporary, increase in radius, effective temperature, and intrinsic flux. The heat interior to its initial outer stabilizing $Z$ gradient is released upon mixing when the model reaches $\sim$0.4 Gyr, at which the outer regions ($\gtrsim 0.6$ M$_P$, where M$_P$ is the planet mass) begin to cool adiabatically. Adiabatic cooling of the outer regions leads to greater internal flux and significantly smaller radii, as shown in the center and center bottom panels of Figure~\ref{fig:fig3}. In contrast, maintaining the outer compositional gradient results in significantly lower luminosities and larger radii. These example models serve illustrative purposes only, highlighting the impacts of varied initial entropy gradients rather than attempting to precisely reproduce Uranus’s observed properties.

The demonstration models in Figure \ref{fig:fig3} show that a sufficiently high initial interior entropy will mix the envelope, leading to systematically higher luminosities and smaller radii than models which do not mix (also shown in Appendix B, Figure \ref{fig:fig9}). These models also show that to keep a stabilizing outer layer and better match the radius of a Uranus-mass model, its interior entropy (for this ice mixture ratio) should be $\sim$3 k$_B$ baryon$^{-1}$. This corresponds to initial interior temperatures of $\sim$15,000 K. When the heavy constituents are restricted to pure water mixed with H–He, comparable results are obtained with an entropy of roughly 2.4 k$_B$ baryon$^{-1}$ (See Appendix A).

Figure \ref{fig:fig3} further shows that entropies of $\sim$3.8 k$_B$ baryon$^{-1}$ undermine the stabilizing $Z$ gradient, and larger values do so at progressively earlier ages. Models that preserve the stable layer (green curve) emit higher luminosities than those that do not (red curve), yet they predict noticeably smaller radii, as seen in the bottom-center panel. The final parameters considered in this study are visualized as a function of initial interior entropy in Appendix B, Figure \ref{fig:fig9}. Entropies $\gtrsim$~3.9 k$_B$ baryon$^{-1}$ fully homogenize the interior. For a pure-water EOS with H–He, destabilization requires about 3 k$_B$ baryon$^{-1}$. Mixtures with lower methane–ammonia content thus need entropies between these bounds, whereas mixtures richer in methane and ammonia demand even higher values to erase the outer compositional gradient.

\section{Evolution of Uranus and Neptune}\label{sec:evolution}

Section \ref{sec:experiments} demonstrated that initial envelope entropies at $\sim$3.8 k$_B$ baryon$^{-1}$ can destabilize the outer composition gradient.\footnote{This interior entropy value applies only to the ice mixture ratio being used here. This entropy value is different for different compositions. The goal of this work is not to firmly constrain these numbers since they are subject to all the uncertainties related to Uranus and Neptune.} This is a similar idea to the results of VH20, which preserved an outer composition gradient in a Uranus model to keep the interior heat within the planet and allow the outer regions to cool. In contrast, Neptune's higher luminosity suggests that its outer composition gradient requires more substantial internal energy replenishment. To reduce the convective stability of Neptune’s outer compositional gradient, we assigned interior entropy profiles higher than Uranus for a particular $Z$ and rock mass fraction distribution. Recall from Section \ref{sec:experiments} that models immediately above the convective mixing threshold satisfy two requirements demanded by the present-day Neptune at once$-$ a higher present-day intrinsic flux and a smaller present-day radius compared to Uranus. As such, our standard Neptune initial conditions require higher initial entropies than Uranus by $\sim$0.5--1 k$_B$ baryon$^{-1}$ to destabilize the outer region, depending on the rock mass distribution. Values higher than this range can completely destabilize the profile, leading to a homogeneous and adiabatic evolution, which does not yield the expected gravity data of Neptune (and even smaller radii). These entropy constraints lead to a higher initial interior temperature of Neptune of between 2,000--5,000 K relative to initial interior temperatures of Uranus.

We performed a parameter search varying the compact rocky core mass, initial outer metallicity, envelope rock mass fraction, and moment of inertia factor ($I/MR^2$), as summarized in Table \ref{tab:tbl2}. Previous investigations identified Uranus models with core masses below 1 M$_\oplus$ and moment of inertia factors of around 0.222 \citep[e.g.,][]{Nettelmann2012, VazanHelled2020}, whereas VH22 formation models suggested more substantial cores of about 4 M$_\oplus$, motivating the range of the parameter search. Neptune models with compact core masses of up to 3 M$_\oplus$ have been compatible with previous models \citep{Nettelmann2013a}. Interior modeling studies have previously permitted outer metallicities as low as 1$Z_\odot$ \citep{Nettelmann2012, VazanHelled2020} for Uranus, despite \textit{Voyager 2} observations indicating Uranus’s atmospheric metallicity to be at least 30$Z_\odot$ \citep[corresponding to $Z \gtrsim 0.34$; see Figure 2 of][and references therein]{HelledHoward2024}. Additionally, although VH20 explored interior rock fractions ranging from one-third to two-thirds by mass, higher ice-to-rock ratios remain plausible \citep{Bethkenhagen2017}. Therefore, our parameter range includes rock mass fractions from 0.05 up to 0.35.

All models are evolved to 5 Gyr and their present-age parameters compared with observations (Table \ref{tab:params}). These models are scored and ranked according to their fractional errors with respect to the observations, so the models with the least fractional errors are presented here. An example visualization of the relative fractional errors considered here is shown and discussed in Appendix B, Figure \ref{fig:fig10}. Figure \ref{fig:fig4} showcases the evolution of temperature (top panels) and density (bottom panels) profiles for preferred models of Uranus (left panels) and Neptune (right panels). The selected initial conditions yield a present-day Uranus model consistent with its observed radius, effective temperature, intrinsic thermal flux, and gravity harmonics, as depicted in Figure \ref{fig:fig5}. Due to convective stability imparted by the outer compositional gradient, the outer layers of Uranus (M$ > 0.92$ M$_P$) cool to temperatures under 1000 K. The interior temperatures stabilize between approximately 6,500 K and 16,000 K and do not cool significantly. This is consistent with the Uranus models of VH20. The top panel of Figure \ref{fig:fig4} compares these interior temperatures to the phase boundaries of water, ammonia, and methane, showing that Uranus’s interior remains within their supercritical phases. At approximately 0.8 Gyr, Uranus’s temperature profile crosses below the hydrogen-water miscibility curve of \cite{Gupta2024} (blue dotted line), suggesting potential water phase separation. The compact rocky core, characterized by $Z = 1$, is approximately 0.5 M$_\oplus$. The rock mass fraction throughout the inner envelope is 0.35, resulting in an overall ice-to-rock ratio of about 1.5. The moment of inertia factor of the Uranus model is 0.2222, in agreement with the past work of \cite{Nettelmann2013a} and VH20.

Recall from the previous section that higher initial interior entropies systematically yield higher internal fluxes and smaller radii. A Neptune model must show a higher luminosity and smaller radius relative to the Uranus model. As a result, higher initial entropies than Uranus were chosen for the Neptune models (motivated by the models in Figure \ref{fig:fig3}). This higher initial entropy leads to convective mixing at 0.5 Gyr, triggering a sudden release of interior heat to Neptune’s outer regions. As depicted in the right column of Figure \ref{fig:fig5}, this rapid heat release transiently elevates Neptune's temperature and intrinsic flux, analogous to the hot model depicted in red in Figure \ref{fig:fig3}. Following this brief episode, the outer regions ($\gtrsim$ 0.4 M$_P$) of the structure cool adiabatically and the radius swiftly contracts, reaching the expected intrinsic flux, effective temperature, and radius at the current age of 4.56 Gyr. At this stage, Neptune's outer envelope attains temperatures of 10,000 K at 30\% of its interior mass and 2,000 K at 60\%. In comparison, the corresponding mass coordinates in the Uranus model maintain temperatures of approximately 13,000 K and 8,000 K, respectively. Such internal temperatures in Neptune's envelope could permit ice phase transitions at later evolutionary stages, as illustrated in Figure \ref{fig:fig4} (right column). Consequently, portions of Neptune's envelope might undergo a supercritical-to-ice phase transition. Additionally, Neptune may experience hydrogen-water phase separation early in its evolution, since it cools to temperatures below the hydrogen-water miscibility curve from \cite{Gupta2024}. The cooler envelope temperatures at later ages suggest a larger fraction of Neptune’s envelope may be undergoing water depletion. The envelope rock mass fraction in the presented Neptune model is 5\% by mass, with a core mass of 3.5 M$_\oplus$. The moment of inertia factor of the Neptune model is 0.225.

Figure \ref{fig:fig5} indicates that the primary evolutionary changes in the Uranus model occur within the initial 0.6 Gyr. The internal flux luminosity of the model aligns closely with the observational data from \cite{Pearl1991} and the more recent measurements by \cite{Irwin2025} and \cite{Wang2025}. Notably, this model accurately reproduces the refined gravity harmonic measurements reported by \cite{Jacobson2025}, which are informed by updated orbital data of Uranus's rings and moons from \cite{French2024}. The updated  measurement from \cite{Jacobson2025} represents an order-of-magnitude improvement in precision compared to \cite{Jacobson2014}. The Uranus model in Figures \ref{fig:fig4} and \ref{fig:fig5} differs by 0.1\% and 0.4\% from the observed $J_2$ and $J_4$ values, respectively. Readers should note that the measured uncertainties of the Uranian $J_2$ are $\sim$0.002\%. Evolution models carry more inherent uncertainties than present-day static models \citep[e.g.,][]{Movshovitz2022, Neuenschwander2022, Lin2024} whose aim is to match the gravitational harmonic data as close as possible.\footnote{For instance, evolutionary models of Jupiter \citep[e.g.,][]{Vazan2018, muller2020, Tejada2025, Sur2025a} cannot be evaluated against present-day criteria with the same rigor as static interior models \citep[e.g.,][]{Wahl2017, Nettelmann2021, Militzer2022, Militzer2024}, due to the additional complexities involved in modeling planetary evolution.} Unless a more complete physical understanding of a planet's interior is obtained, evolution models will be more inherently uncertain. Given the absence of substantial convective mixing in Uranus, its outer heavy element abundance, represented by the dashed line in the top row of Figure \ref{fig:fig5}, remains relatively stable throughout its evolution.

As for the Neptune model, we compare our models against intrinsic flux measurements from \textit{Voyager 2} calculated by \cite{Pearl1991}, and gravity harmonic measurements by \cite{Jacobson2009}. Due to the Neptune model's rapid convective mixing event, the outer heavy element abundance rises quickly from an initial value of 25\% to $\sim$80\% by mass. The Neptune model reproduces the observed $J_2$ and $J_4$ values compared to the median measurements of \cite{Jacobson2009} are 0.8\% and 0.3\% respectively. A comprehensive comparison between modeled and observed values of both planets is tabulated in Table \ref{tab:results}. 

Inferring the density profiles of Uranus and Neptune has been the subject of extensive research as a way to constrain the interior structure, mainly by varying the density profiles to match the gravity harmonic measurements \citep[e.g.,][]{Podolak1981, Podolak1991, Podolak1995, Marley1995, Helled2011, Movshovitz2022, Lin2024}. \cite{Movshovitz2022} computed a baseline of density models for Uranus and Neptune based on the available gravitational harmonic data. The outline of their family of density models ranges from $\sim$2--17 g cm$^{-2}$ (see their Figure 3). The reader should note that the baseline region of \cite{Movshovitz2022} represents the envelope of possible density profiles, not the density profiles themselves. The Uranus and Neptune models in Section \ref{sec:evolution} at 4.56 Gyr are comfortably consistent with their baseline values. Our Uranus and Neptune models are less dense than the 3-layer models of \cite{Nettelmann2013a} due to their higher internal temperatures and ice mixture compositions. Figure \ref{fig:fig6} shows the model density profiles of Uranus (top) and Neptune (bottom) at 4.56 Gyr compared to other models in the literature \citep{Nettelmann2013a, VazanHelled2020, Neuenschwander2022, Neuenschwander2024, Morf2024, Malamud2024}. Of the density profiles shown in Figure \ref{fig:fig6}, only those of \cite{Nettelmann2013a} and \cite{VazanHelled2020} are evolution models.   

The evolution models in this section are the first inhomogeneous, non-adiabatic models of Uranus and Neptune which incorporate formation-informed initial conditions, including methane-ammonia-water-rock ice mixtures throughout the envelopes of both planets. In addition, these evolution models are the first to be informed by the recent empirical work of \cite{Lin2024}, which suggests that the interiors of both planets may include higher amounts of H-He-ice mixtures to better explain the gravity harmonics. The Uranus and Neptune models shown in Figures \ref{fig:fig4} and \ref{fig:fig5} are also the first to agree with all the observable parameters in Table \ref{tab:results} simultaneously within reasonable uncertainty. Particularly, the Uranus model is well within the range of the updated internal flux measurements of \cite{Irwin2025} and \cite{Wang2025}, as well as the updated gravity measurements of \cite{Jacobson2025}.

\section{Conclusion}\label{sec:conclusion}

Formation models indicate that both Uranus and Neptune could have initially possessed substantial outer compositional gradients \citep{Valletta2022}. To account for Uranus's low internal luminosity, such a gradient likely restricts heat transfer from the interior, maintaining a cooler exterior \citep{Nettelmann2016, VazanHelled2020, Scheibe2021}. Neptune’s higher luminosity could be explained by convective mixing of its outer compositional gradient formed during its early evolution. Past work on inhomogeneous and non-adiabatic gas giant evolution has shown that lower initial entropies in an entropy gradient better preserve stable regions over evolutionary timescales, while stable regions located farther out the structure tend to convectively mix stable regions \citep{Knierim2024, Tejada2025}. These two observations have been applied here. Thus, we conclude that the formation models of VH22 can explain the present-day Uranus and Neptune luminosity differences, which could result from variations in their initial entropy/thermal profiles and subsequent heavy element profile evolution. Importantly, we propose that convective mixing could be the primary reason for Neptune's high internal heat flux whenever the initial interior entropy destabilizes the outer stable region over evolutionary timescales.

Our supplementary conclusions are as follows:

\begin{enumerate}
    \item In agreement with VH20, the interior regions of Uranus  ($\gtrsim 1$ kbar) have likely remained hot and supercritical since formation, provided its outer composition gradient remains convectively stable. The requirement of long-term convective stability is a consequence of 1) an expected non-adiabatic and inhomogeneous interior \citep{Scheibe2019, Neuenschwander2022, Neuenschwander2024, Lin2024} and 2) the need to explain its low measured internal flux \citep{Nettelmann2016, VazanHelled2020}.
    \item Provided that substantial regions of Neptune’s interior ($\gtrsim$0.6 M$_P$) undergo post-mixing adiabatic cooling, ices in these regions may undergo a supercritical-to-solid phase transition. 
    \item Both planets may experience H$_2$/water phase separation, with Neptune potentially undergoing more significant water depletion. 
    \item If Neptune experiences more convective mixing, as suggested by this work, Neptune may have higher atmospheric metallicity than Uranus.
\end{enumerate}

The prediction that Neptune may have a higher atmospheric metallicity than Uranus is supported by atmospheric measurements \citep[e.g., Figures 7 and 2 of ][respectively]{Guillot2023, HelledHoward2024}. The atmospheric metallicity measurements of Uranus and Neptune range between $\sim$30--74 and $\sim$40--93 times the solar abundances respectively.\footnote{Ranges are based on C/H and S/H ALMA, VLA, HST, and Keck measurements listed in Table 2 of \cite{Guillot2023}.} In addition to atmospheric measurements, interior models predict that Neptune's outer metallicity could be higher than that of Uranus. For example, early work by \cite{Podolak1995} predicted that the atmospheric composition of Neptune would be very different from that of Uranus. Guided only by static interior density models and early gravity data, \cite{Podolak1995} could not explain this difference at the time. Another example is the models of \cite{Nettelmann2013a}, whose Neptune models show a greater outer heavy element abundance than their Uranus models. Super-adiabatic atmospheric temperature profiles studied by \cite{Leconte2017} and \cite{Cavalie2017} show that the O/H ratio of Uranus should be less than 160 times the solar value, and about 540 times solar in Neptune. 

The geometric albedo spectrum features of Neptune are deeper than that of Uranus at shorter wavelengths \citep{Karkoschka1994, Karkoschka1998}. The early work of \cite{Karkoschka1998} found that both Uranus and Neptune show a greater abundance of methane in their atmospheres compared to Jupiter and Saturn, but methane features in Neptune are more pronounced than in Uranus. Since geometric albedo is a weak function of abundance \citep[][]{Madhusudhan2012, Burrows2014}, small differences in  spectral depth indicate greater abundance differences. Hence, the differences between Uranus and Neptune at shorter wavelengths imply a greater methane abundance in Neptune's atmosphere.

\begin{figure*}[ht!]
\centering
\includegraphics[width=0.47\textwidth]{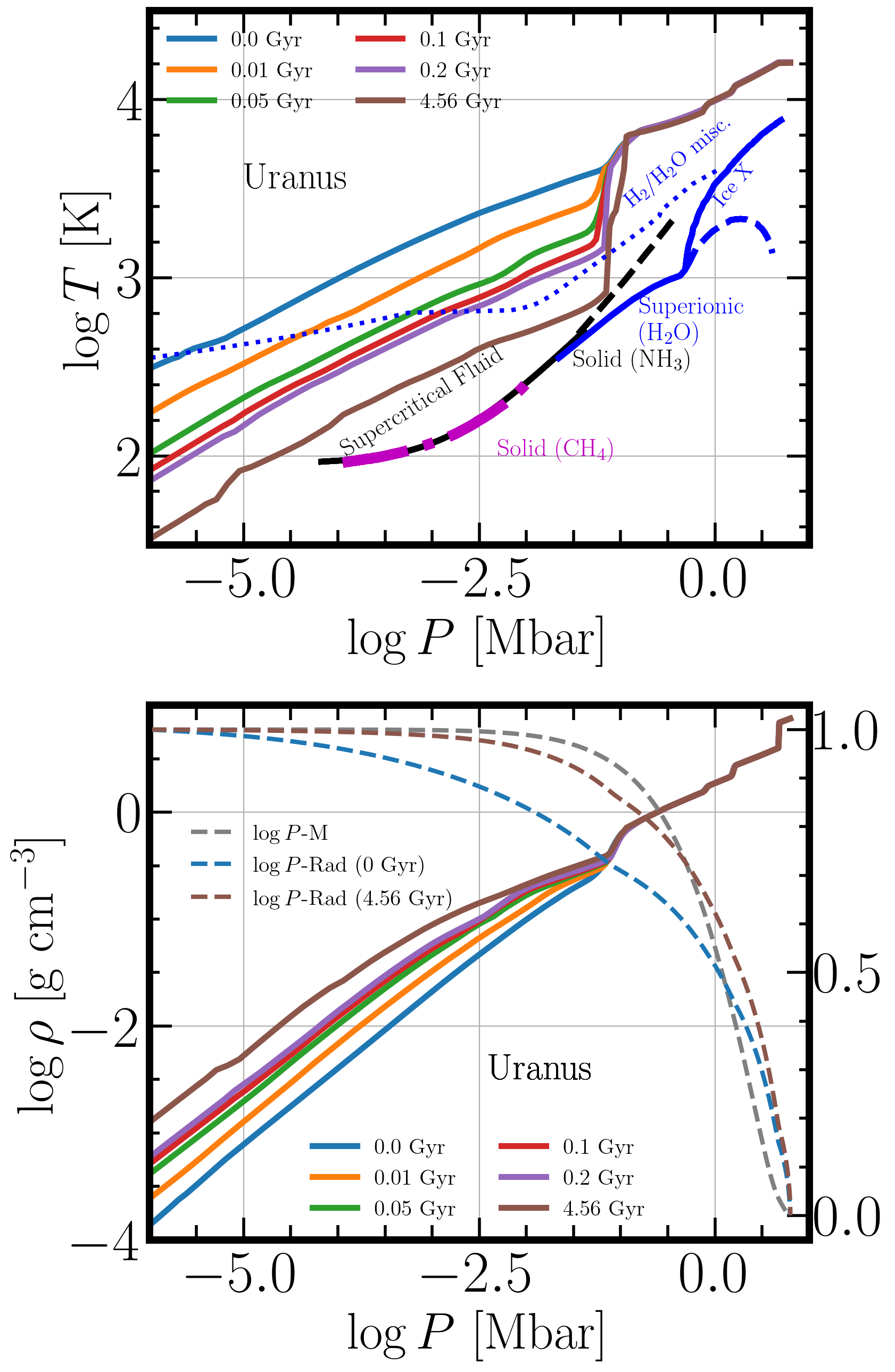}
\includegraphics[width=0.47\textwidth]{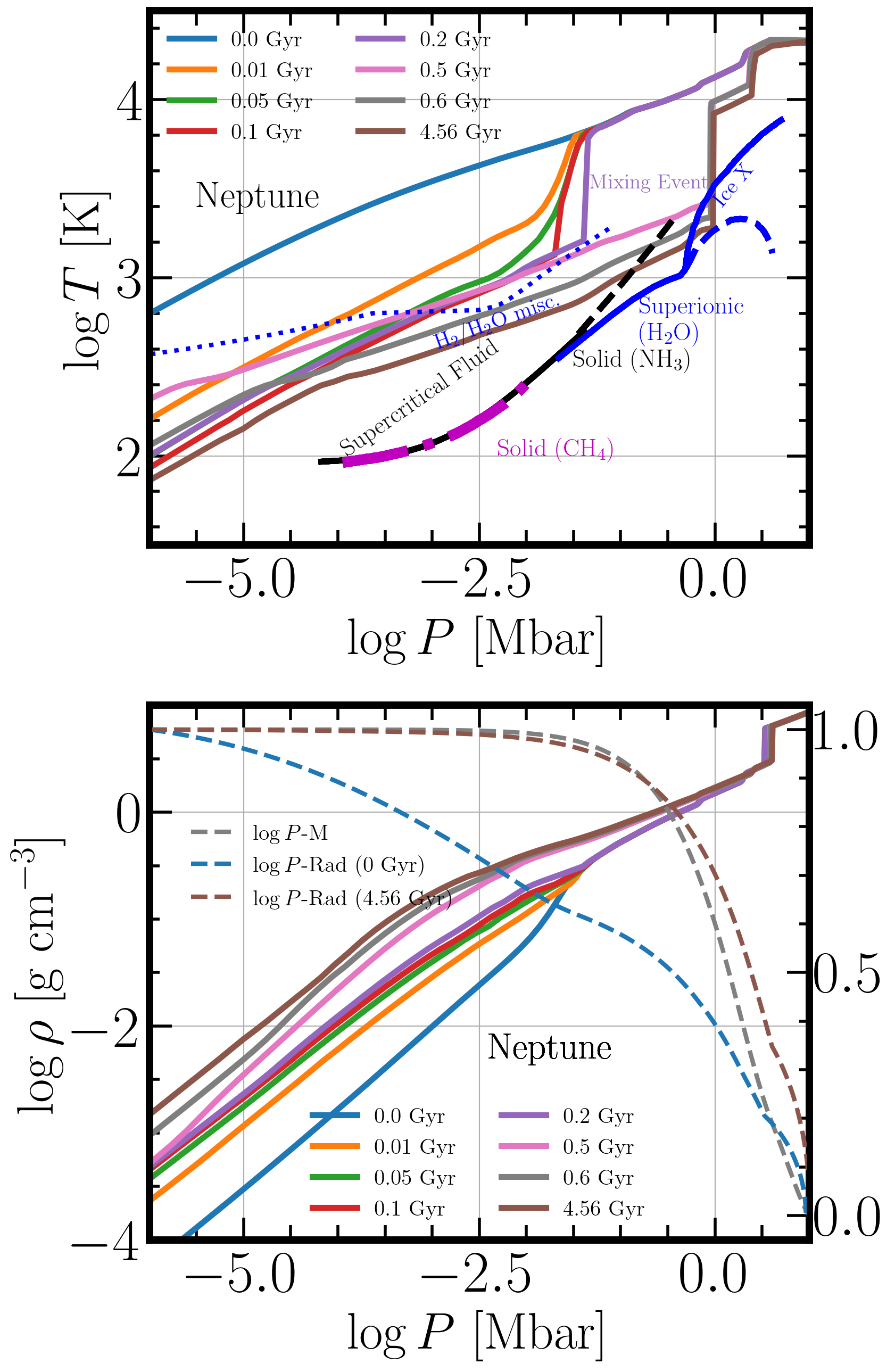}
\caption{Evolution of the temperature (top row) and density profiles (bottom row) of Uranus (left) and Neptune (right) as a function of pressure. Top: The outer regions of Uranus ($\lesssim 1$ kbar, 0.92 M$_P$) cool, while the interior maintains its initial heat. As a result, the interior temperature does not escape the outer boundary layer. In contrast, Neptune experiences more vigorous convective mixing, leading to the outer boundary layer mixing after 0.5 Gyr. After this mixing event (labeled in the top right), the outer $40\%$ of Neptune's structure cools adiabatically. Along side the profiles, the phase diagram boundaries between the supercritical and solid states for methane, ammonia, and water are plotted. The water phase boundaries were gathered from \cite{Millot2018}. The blue dotted line indicates the miscibility curve at which H$_2$ and water are predicted to phase separate \citep{Gupta2024}. The miscibility curves are adapted to the local thermodynamic conditions (pressure, temperature, $Z$) of each structure at 4.56 Gyr. The black dashed melt curve of ammonia was gathered by \cite{Queyroux2019} and extended to low pressures via the Simon-Glatzel melting curve fit for ammonia \citep{Ninet2008}, shown in solid black. The methane melting curve is from \cite{Somayazulu1996} as cited by \cite{Bini1997}. The interior heat of Uranus escapes inefficiently via conduction through the outer composition gradient. On the other hand, Neptune's outer layers convectively mix and cool to temperatures below the freezing temperatures and pressures of water at the Ice X phase. Both planets intercept the H$_2$-H$_2$O miscibility curve of \cite{Gupta2024}, implying that water and hydrogen may phase separate in both planets. Bottom: The density profile evolution of Uranus changes only in the outer regions. Once Neptune's outer $Z$ gradient mixes, the interior increases in density as it cools adiabatically. The gray dashed line indicates mass as a function of pressure. The blue and brown lines indicate normalized radius as a function of pressure at the corresponding color age (0 and 4.56 Gyrs).}
\label{fig:fig4}
\end{figure*}

\begin{figure*}[ht!]
\centering
\includegraphics[width=0.47\textwidth]{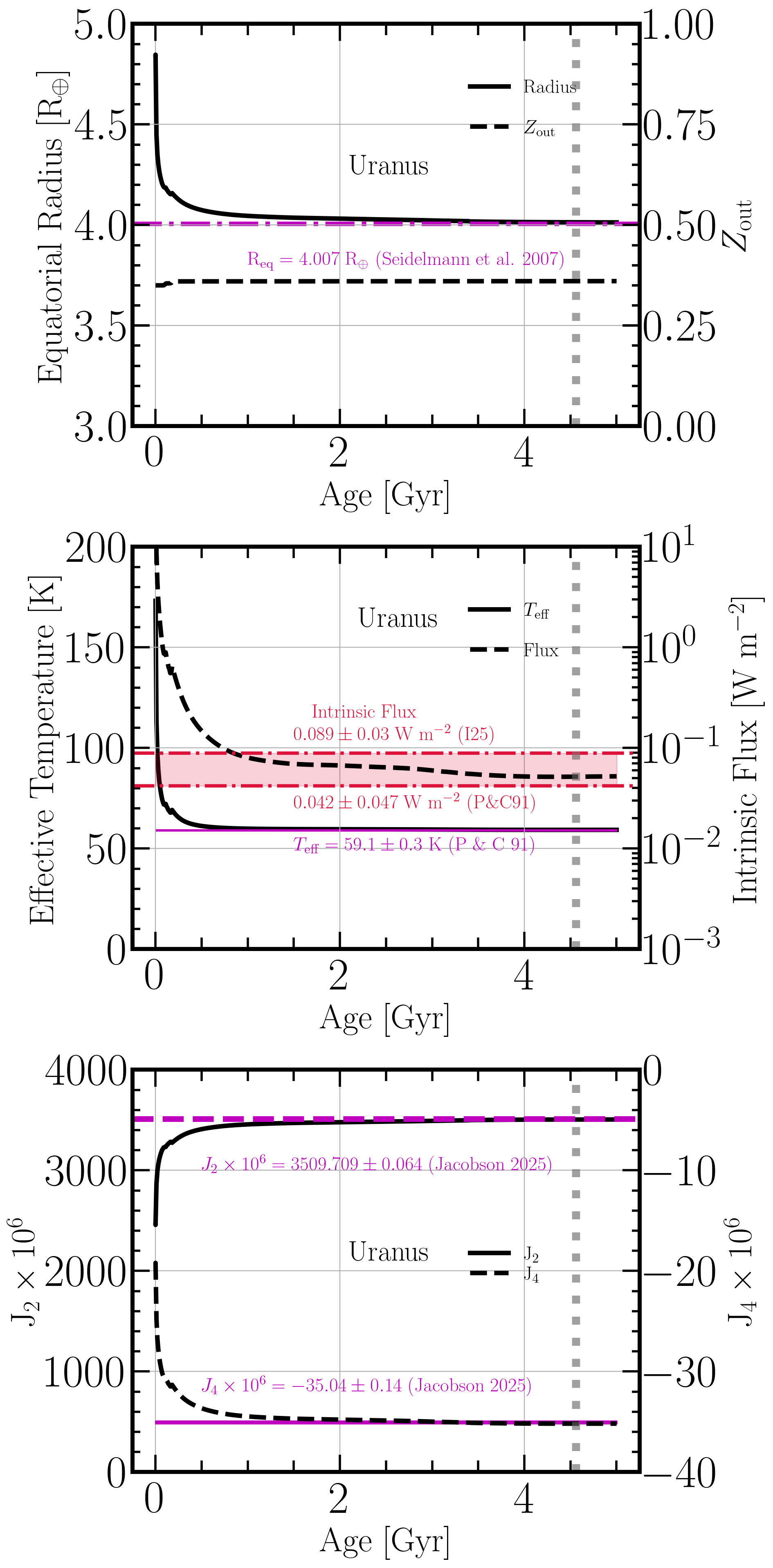}
\includegraphics[width=0.47\textwidth]{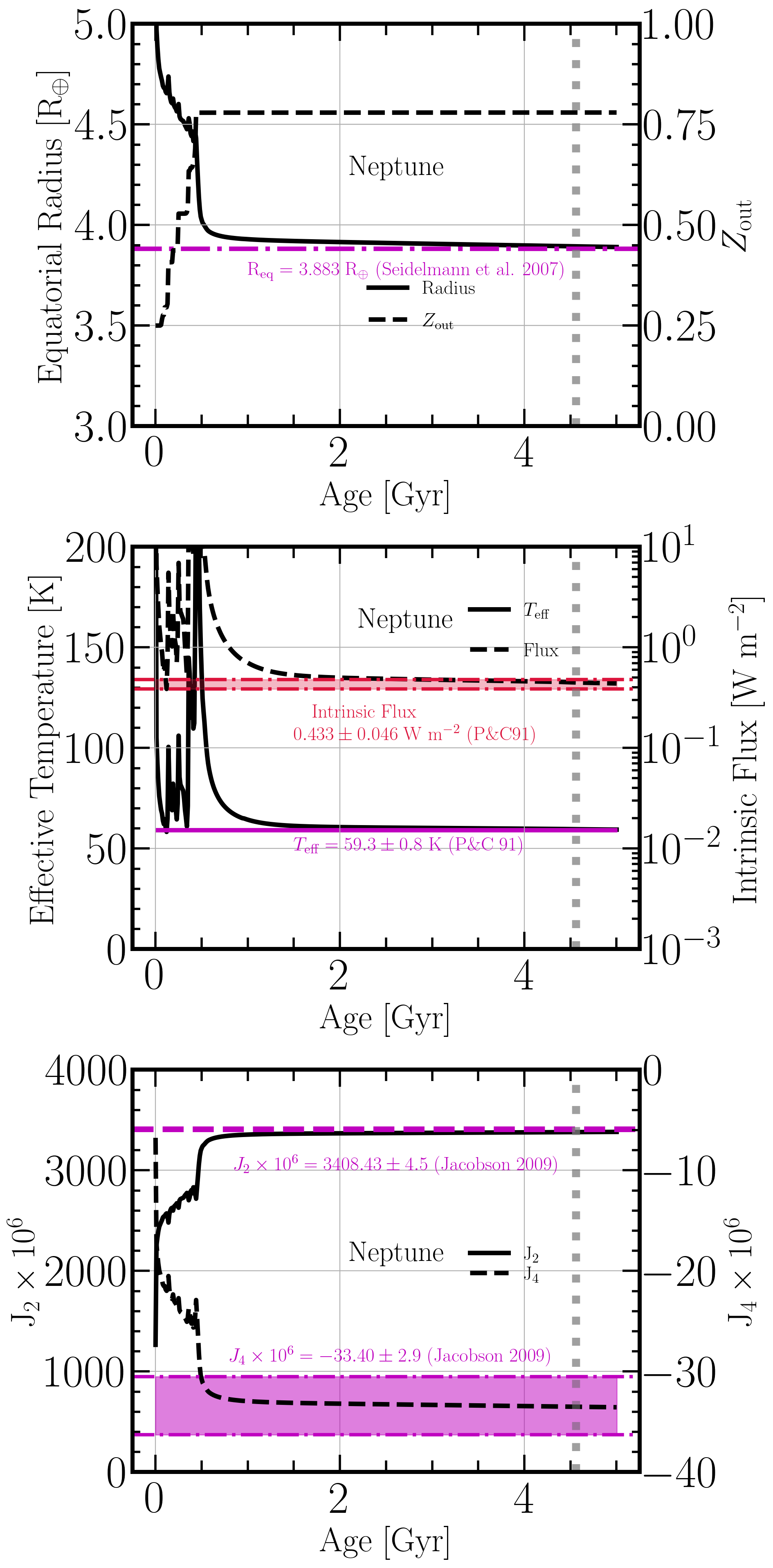}
\caption{Evolution of various quantities of the Uranus (left) and Neptune (right) models shown in Figure \ref{fig:fig4}. The evolution of the radius and the outer $Z$ abundance is shown in the top row, the effective temperature and intrinsic flux in the middle row, and gravity harmonics in the bottom row. The initial outer $Z$ abundances of the Uranus and Neptune models are 0.35 and 0.25 by mass and the final abundances are 0.36 and 0.78 by mass. The initial metallicities of these models are the result of the parameter search described in Section \ref{sec:evolution}. The internal luminosity of Uranus considered here is between the medians of \cite{Pearl1991} and \cite{Irwin2025}, shown in red. There are no updated luminosity measurements of Neptune since those measured by \cite{Pearl1991}. The 1 and 2 sigma errors of the intrinsic flux from \cite{Pearl1991} are shown in red and gray in the center-right panel. The model gravity harmonics of Uranus are compared with the updated $J_2$ and $J_4$ measurements of \cite{Jacobson2025} since the uncertainties of their $J_4$ measurements are an order of magnitude smaller than those of \cite{Jacobson2014}. The Neptunian $J_4$ uncertainties from \cite{Jacobson2009} are significant. The Uranus model evolves to within 10\% of the current Uranus observations within 0.6 Gyr, while the outer composition profile of the Neptune model convectively mixes at $\sim$0.5 Gyr. The sudden mixing in Neptune's envelope leads to the abrupt release of internal energy, heating up the envelope. Once heated, the outer envelope cools adiabatically. This access to internal heat post-mixing lends Neptune a luminosity boost, keeping its luminosity higher relative to Uranus.  }
\label{fig:fig5}
\end{figure*}

\begin{figure}[ht!]
\centering
\includegraphics[width=0.47\textwidth]{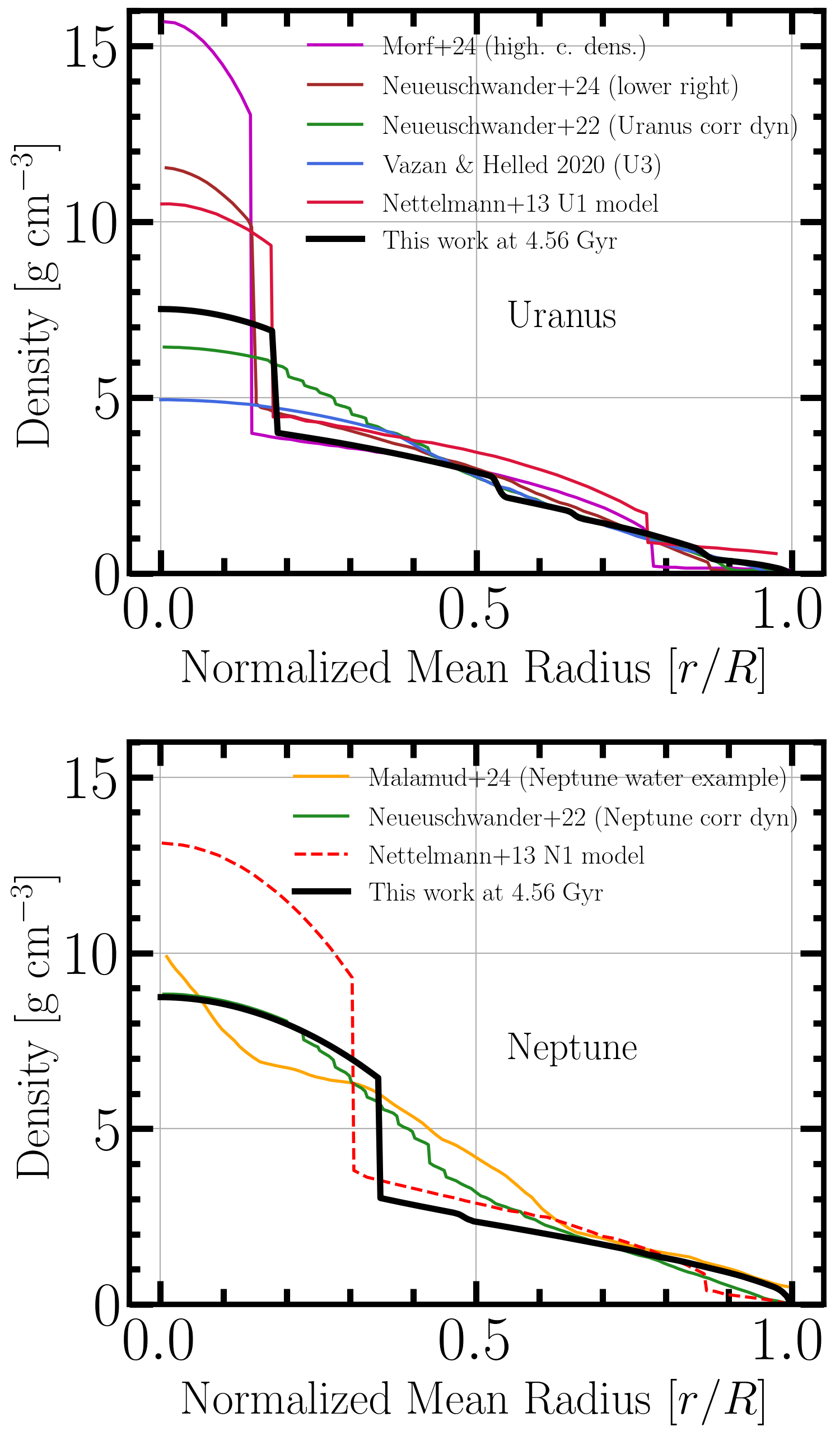}
\caption{Present-age density profiles of the Uranus and Neptune models (thick black lines) presented in Section \ref{sec:evolution} compared with models from the literature \citep[color lines;][]{Nettelmann2013a, VazanHelled2020, Neuenschwander2022, Neuenschwander2024, Malamud2024}. Of these, only the Uranus models of \cite{Nettelmann2013a} and \cite{VazanHelled2020} are the result of evolutionary calculations. Others are static inferences. Since these authors produce multiple models with varying of water/rock compositions, we indicate which are plotted in parentheses. Our Uranus models are consistent with literature results. There are fewer Neptune density profiles in the literature, but ours is consistent with the Neptune model of \cite{Nettelmann2013a}. \cite{Neuenschwander2022} produced their models using so-called ``empirical'' EOSes, while \cite{Malamud2024} showed only an example profile instead of a preferred model. The red dashed density profiles are those of \citet[][N13]{Nettelmann2013a}. Our models are overall less dense than the 3-layer models of N13 due to their higher interior temperatures and ice mixtures.}
\label{fig:fig6}
\end{figure}

We emphasize that the ``ices'' within Uranus and Neptune may deviate from a strictly solar composition if they experience disassociation and phase separation \citep{Cheng2021, Cheng2023, Militzer2024b}. Nevertheless, the interior temperatures of these planets may not be sufficiently low to promote the formation of distinct layered structures. Complete phase separation into distinct ice layers likely requires colder temperatures, potentially associated with superionic phase transitions \citep{Millot2018}, as illustrated by the example static 3-layer profiles shown by \cite{Militzer2024b}, gathered from \cite{Redmer2011}, to propose the phase separation of ammonia and methane from water. Past evolution models \citep{VazanHelled2020} and recent empirical gravitational harmonic analysis \citep{Lin2024} support the persistence of supercritical ice-rock mixtures at higher temperatures, consistent with the findings of this study. Thus, the interior conditions of Uranus may remain too warm for substantial layering and core freezing, as suggested by \cite{Stixrude2021}. However, Figure \ref{fig:fig4} indicates that supercritical-to-solid ice phase transitions of water, ammonia, and methane could occur in the outer envelope of Neptune. We highlight that the specific entropy quantities required to destabilize any given stable layer depend on the scale of the stabilizing entropy gradient (see end of Section \ref{sec:experiments}), which depends on the heavy element equation of state. Models with different volatile abundances relative to water affect the initial conditions and final state of a planet significantly (see Appendix A, Figure \ref{fig:fig8}). However, the general physical idea that a sufficiently high initial interior entropy (i.e., a sufficiently low entropy gradient) convectively mixes an initial $Z$ profile gradient over evolutionary timescales \textit{is independent of ice mixture ratios}. The Uranus mass models with different ice mixtures in Figure \ref{fig:fig8} could, in principle, be brought to match the radius and gravity data if initialized at higher entropies, and could convective mix if initialized at higher initial entropies. A future systematic study on the dependence on volatile ratios, rock mass fractions, and other parameters is surely warranted. Therefore, the general main result of this work holds for any mixture of ices and rocks mixed with hydrogen and helium.

A possible scenario which could lead to a higher initial interior entropy for Neptune is a giant impact. Giant impacts during the formation of Neptune are a potential explanation for its hotter initial entropy. Such early collisions could modify the thermal and compositional structures of both Uranus and Neptune \citep{Reinhardt2020}. A sufficiently oblique impact can reproduce Uranus's extreme obliquity and its prograde, coplanar satellite system \citep{Morbidelli2012}, yet subsequent simulations suggest that an impact alone may not fully account for the present satellite orbits \citep{Rufu2022}. The work of \cite{Reinhardt2020} predicts that a giant impact could be enough to tilt Uranus's spin axis and produce an aligned debris disk. Those authors also conclude that a giant impact could render the Neptunian interior adiabatic. The influence of a giant impact on the interior of Uranus, however, is less clear \citep{Reinhardt2020}. Elsewhere in the Solar System, giant impacts have also been invoked to explain Jupiter's extended core, but recent hydrodynamical studies indicate that even head-on collisions cannot create the observed fuzzy core (i.e., a stable region) \citep{Sandnes2024, Meier2025}. A giant impact on Neptune early in its evolution is a compatible hypothesis not only to explain its higher initial entropy, but also its higher mass rocky core.

Another process that affects gas giant planet evolution is H-He phase separation \citep{Stevenson1975, FortneyHubbard2003, FortneyHubbard2004, Mankovich2016, Mankovich2020, Tejada2025, Sur2025a}. Instead of H-He separation, water-hydrogen phase separation is expected in both planets (See Figure \ref{fig:fig4}). Future studies will incorporate water and hydrogen phase separation, potentially impacting structural and magnetic evolution \citep{Bailey2021, Gupta2024, Amoros2024}. Hydrogen-water immiscibility, particularly in outer H-He-rich regions at pressures of 10–30 GPa and temperatures below $\sim$6,000 K \citep{Gupta2024, Amoros2024}, could influence planetary luminosity and composition profile evolution. Notably, water may separate independently from ammonia and methane, which themselves may become immiscible with water at lower temperatures \citep{Militzer2024b}. This work suggests that if hydrogen-water phase separation occurs (or even ice phase separation), then the interior temperatures of Neptune may provide more favorable conditions. 

Formalizing and expanding the parameter search outlined in Section \ref{sec:evolution} is a promising topic for future work. The space of possible initial conditions is vast, owing to the many combinations of ice–rock mixtures and the stabilizing entropies they require. Here we show that convective mixing can plausibly account for the divergent evolutions of Uranus and Neptune, and that the relevant initial conditions are, in principle, \textit{inferrable} once their interior compositions are better constrained. A similar strategy has already been applied to gas-giant exoplanets \citep{Knierim2025}. A comprehensive inference analysis for Uranus and Neptune is therefore desirable, but its reliability will depend on forthcoming advances in mixture equations of state and in key microphysical inputs such as ice-mixture conductivities and diffusion coefficients.

In this work, we used analytical atmospheric fits from \cite{Graboske1975} and \cite{Hubbard1977} as presented by \cite{Guillot1995}. The albedos of both planets are uncertain and updated albedos are necessary to better model the atmospheric cooling. An updated Uranian albedo of 0.349 has been published recently by \cite{Irwin2025}, but this difference of albedo \citep[0.3 from \textit{Voyager 2};][]{Pearl1990a} leads to negligible differences in Uranus evolution models. Past work \citep[e.g.,][]{Mankovich2020, Chen2023} has shown that a different atmospheric albedo from 0.3 to 0.5 significantly affects evolutionary results for Jupiter. Self-consistent irradiation models with clouds and up-to-date opacities can make a significant difference ($\gtrsim 6\%$) in the evolution of Jupiter and Saturn \citep{Chen2023}. Particularly, the presence of clouds introduces condensation features which result in deviations from models without clouds \citep[][]{Sudarsky2000, Chen2023}. Since clouds are expected in the troposphere of both planets \citep{Gautier1995}, clouds could, hence, affect the thermal evolution of Uranus and Neptune. Future research incorporating more sophisticated atmospheric treatments$-$ including solar irradiation, high-metallicity opacities, and the effects of clouds \citep{Chen2023, Tejada2024, Sur2025a}, is needed. Importantly, better boundary conditions in the presence of convective mixing may slow down the rate of cooling of Neptune instead of abruptly heating up the outer layers. This effect could be more pronounced at low temperatures where hazes and condensates could increase the geometric albedos \citep{Fortney2011}. Evolutionary models with irradiation and metallicity-dependent atmospheric boundary conditions with the effects of clouds will be featured in future work (Tejada Arevalo et al. 2025c, \textit{in prep}).

Heat conduction governs energy transport across convectively stable layers when semiconvection (double-diffusive convection) is absent. In this study, we combine the water thermal conductivities of \cite{French2019} with the H-He conductivities described in \cite{Sur2024a}. Thermal conductivity data for ice–rock mixtures are needed and will be essential for refining heat-flux estimates at compositional interfaces. Should semiconvection occur, reliable values for conductivities, Nusselt numbers, and Rayleigh numbers under the conditions expected in Uranus and Neptune are also required. \cite{French2019a} evaluated some of these parameters and concluded that double-diffusive convection is unlikely in either planet. Consequently, composition transport across stable layers is probably purely diffusive, allowing much of each planet’s primordial heat to remain trapped. Our results are consistent with this picture.

Uranus and Neptune remain the least understood planets of our Solar System. Understanding their long-term evolution could inform future formation models, further clarifying feasible formation pathways. Future work will demand a more advanced understanding of ice mixture and ice-rock mixture equations of state, self-diffusion coefficients of ice mixtures, and ice mixture thermal conductivities at various abundances. Aside from better theoretical EOS and microphysical quantities, improved observations of the atmospheric metallicities and gravity harmonics will greatly aid future ice giant modeling \citep[see discussion in][]{Movshovitz2022}. Hence, improvements in mixture EOSes, microphysical transport quantities, and observations have the potential to unlock the history of these elusive planets and also to guide future research towards a better understanding of exoplanets broadly. 

\begin{acknowledgments}
We thank the anonymous referee for their thorough reviews that very much sharpened numerous aspects of the manuscript.
This research was funded by the Center for Matter at Atomic Pressures (CMAP), a National Science Foundation (NSF) Physics Frontier Center under Award PHY-2020249. Any opinions, findings, conclusions, or recommendations expressed herein are those of the authors and do not necessarily reflect NSF views. RTA acknowledges support from the Ford Foundation Predoctoral Fellowship. RTA is grateful to Drs. Adam Burrows, Akash Gupta, Ankan Sur, and Yubo Su for lively discussions, critical reviews and feedback, expertise, and academic mentorship, which significantly influenced this work. RTA is grateful to Dr. Mandy Bethkenhagen for providing equation of state data and valuable discussions. Additionally, RTA thanks Dr. Matthew Coleman and Princeton Research Computing for computational and technical assistance.
\end{acknowledgments}


\begin{deluxetable*}{lcc}
\tablecaption{Searched Parameters for Uranus and Neptune\label{tab:tbl2}}
\tablewidth{0pt}
\tablehead{
  \colhead{Parameter} &
  \colhead{Uranus} &
  \colhead{Neptune}
}
\startdata
Core Mass ($M_{\oplus}$)                         & 0.5--4 & 0.5--4 \\
Initial Outer Metallicity ($Z_{\rm out}$)                 & 0.05--0.45 & 0.05--0.25 \\
Envelope Rock Mass Fraction ($Z_{\rm rocks}$)    & 0--0.35 & 0--0.35 \\
Moment of Inertia Factor         & 0.2--0.23 & 0.2222--0.2555 \\
\enddata
\tablecomments{Quantities varied in the parameter search described in Section \ref{sec:evolution}. The lower and upper bounds of the core mass are informed by the interior model core masses of \cite{Nettelmann2013a, VazanHelled2020}, and the formation models of VH22. Similarly, the moment of inertia factors were informed by the values obtained by \cite{Nettelmann2013a, Nettelmann2016}, and \cite{VazanHelled2020}. The rock mass fractions vary between no rocks and roughly one-third rock composition, similar to models presented in \cite{VazanHelled2020}.}
\end{deluxetable*}

\begin{deluxetable*}{lcccc}
\tablecaption{Uranus and Neptune Evolution Results\label{tab:results}}
\tablewidth{0pt}
\tablehead{
  \colhead{Measurement Quantity} &
  \colhead{Observed Uranus} &
  \colhead{Observed Neptune} &
  \colhead{Uranus Model [4.56 Gyr]} &
  \colhead{Neptune Model [4.56 Gyr]}
}
\startdata
Effective Temperature [K]                         & $59.1 \pm 0.3$ & $59.3 \pm 0.8$& 59.28 & 59.6\\
Intrinsic Flux [W m$^{-2}$]                       &0.042--0.089 & $0.433 \pm 0.046$& 0.052 & 0.448  \\
Equatorial Radius [R$_\oplus$]                    &4.007 & 3.883&4.013 & 3.893 \\
$J_2 \times 10^6$    &  $3509.709 \pm 0.064$ & $3408.43 \pm 4.5$ & 3505.414 & 3380.498\\
$J_4 \times 10^6$         & $-35.04\pm0.140$ & $-33.40 \pm 2.90$ & $-35.18$ & $-33.50$\\
\enddata
\tablecomments{Table of Uranus and Neptune observations used to guide the evolution models in this Section~\ref{sec:evolution}. The observed Uranian $J_2$ and $J_4$ values come from \cite{Jacobson2025} and the Neptunian values from \cite{Jacobson2009}. The tabulated Uranus model quantities are within 0.4\% from the observed values at 4.56 Gyrs. The same values are within 0.6\% for the Neptune model. }
\end{deluxetable*}

\clearpage

\appendix

\section{Appendix A}

\begin{figure*}[ht!]
\centering
\includegraphics[width=\textwidth]{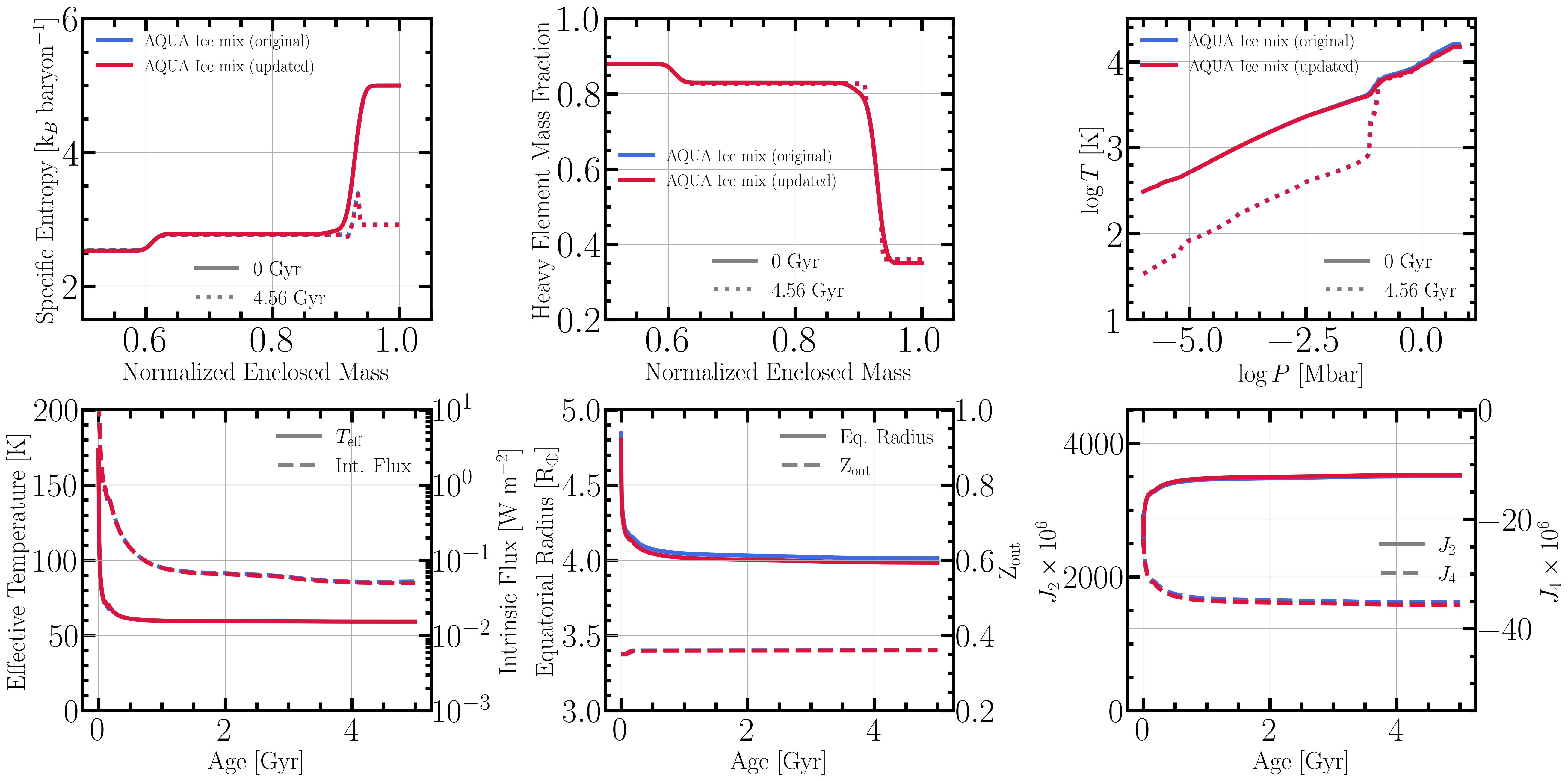}
\caption{Comparison of the Uranus preferred model presented in Section \ref{sec:evolution} and in the left panel of Figure \ref{fig:fig4} with the original and updated AQUA tables. The format of this figure is the same as Figure \ref{fig:fig3}. The same model with H-He-water only is shown in red for example purposes. The updated AQUA table takes into account the entropy correction to the \cite{Mazevet2019} (in magenta) EOS given by \cite{Mazevet2021} (in blue). The differences in the evolution between the original and updated water EOSes are small ($\lesssim 3\%$). The use of the original AQUA EOS does not affect the results presented here. }
\label{fig:fig7}
\end{figure*}

The water EOS used here \citep[AQUA;][]{Haldemann2020} contains the EOS of \cite{Mazevet2019}. As mentioned by \cite{Aguichine2025}, an entropy correction has been issued by \cite{Mazevet2021} which does not affect the density. Figure \ref{fig:fig7} illustrates the differences between the original AQUA and the updated AQUA version incorporating the \cite{Mazevet2021} corrections, tested on the Uranus model presented in Section \ref{sec:evolution}. The evolution of Uranus is not significantly affected by the updated water EOS, and produces qualitatively the same results. We combined the new water EOS of \cite{Mazevet2021} with the original AQUA table  to replace the old \cite{Mazevet2019} EOS and obtain an updated AQUA table. Such a table can be found here: \url{https://github.com/robtejada/eos/tree/main/mazevet}.

\begin{figure*}[ht!]
\centering
\includegraphics[width=\textwidth]{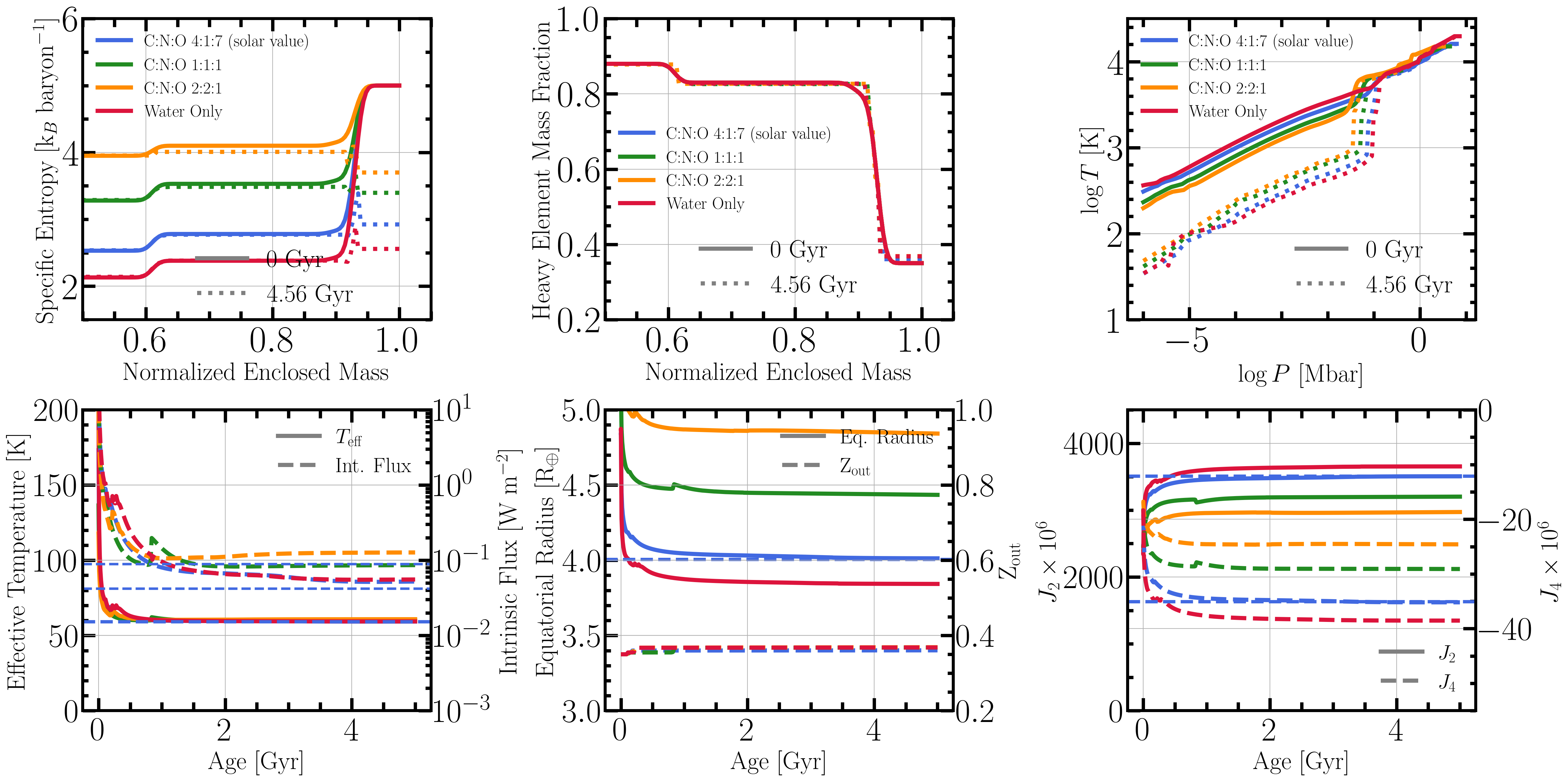}
\caption{Comparison of the evolutionary effects of different ice mixtures on the Uranus model presented in Section \ref{sec:evolution} (shown here in blue). The dashed lines in the bottom panel indicate the measured values to guide the eye. The interior entropy profiles are scaled so that all models begin with similar interior temperatures. All models have a rock mass fraction of 0.35 in their envelopes. Since methane and ammonia are less dense than water, models with more methane and ammonia abundances increase the entropy and thus decrease the density, causing large differences in the total radius and the gravitational harmonics. A model with pure water (shown in red) is more dense and thus exhibits a smaller radius, leading to similar differences in the gravitational harmonic evolution.}
\label{fig:fig8}
\end{figure*}

The chosen ice mixture can have a significant effect on evolution models of Uranus and Neptune. In Figure \ref{fig:fig8}, we compare the Uranus model in Section \ref{sec:evolution} with 1:1:1 and 2:2:1 methane-ammonia-water ratios (we use 4:1:7, shown in blue). These ratios probe mixtures with higher methane/ammonia abundances. We calculate a model with pure water (i.e., no methane or ammonia) to test and compare this extreme case of no methane and ammonia . Generally, lighter materials (e.g., methane/ammonia) exhibit higher entropies at the same temperatures and pressures compared to heavier materials (e.g., pure water, rock). Due to this thermodynamic behavior, the interior entropies of the models were scaled to be initialized so match the interior temperature of the 4:1:7 preferred model. Evolution models with more volatile abundances lead to higher radii and models with low or no volatiles lead to smaller radii relative to the 4:1:7 model. The radius evolution and density profiles can generally be controlled by lowering the compact core mass and controlling the rock mass fractions in the envelope, so models with more volatiles can match the radius with more interior rock abundances and larger compact cores. Conversely, water-only models can match the radius if one uses lower rock abundances and a smaller compact core. Readers should note that, in principle, a broad range of methane-ammonia-water mixtures can be made to match the observable data of both planets by simply controlling the amount of rocks in the envelope, the size of the compact core, and other unknown parameters. We chose the 4:1:7 due to its past use in interior modeling \citep{Bethkenhagen2017} and the precedent to assume solar-like abundances in gas giant planet interiors more generally (e.g., assuming a helium solar abundance of $\sim$0.27). Modeling the change of ices due to phase transitions \citep{Stixrude2021} or immiscibility \citep{Militzer2024b} is left for future work.

\clearpage

\section{Appendix B}

This section further demonstrates the motivation behind our choices of initial entropy and resulting parameters of Uranus-mass models in Sections \ref{sec:experiments} and \ref{sec:evolution}. All exercises in this appendix are repeated for Neptune. Figure \ref{fig:fig8} shows the present-age parameter dependence on the choice of initial internal entropy, for the models in Figure \ref{fig:fig3}. The threshold value of $\sim$3.7 k$_B$ baryon$^{-1}$ is highlighted in gray, and the shaded region to its right represents the adiabatic regime. As we have emphasized in the main text, this threshold value depends on the ice mixture ratios chosen. The broader idea proposed here is that such a threshold exists for any mixture of ices, rocks, or other metals. Evolution models with greater initial interior entropies progressively release more interior heat until this threshold is met, upon which adiabatic evolution takes over (shaded in gray). Models that convectively mix have greater outer metallicities, greater luminosities, and smaller radii. These are all features of the current Neptune.

\begin{figure*}[ht!]
\centering
\includegraphics[width=\textwidth]{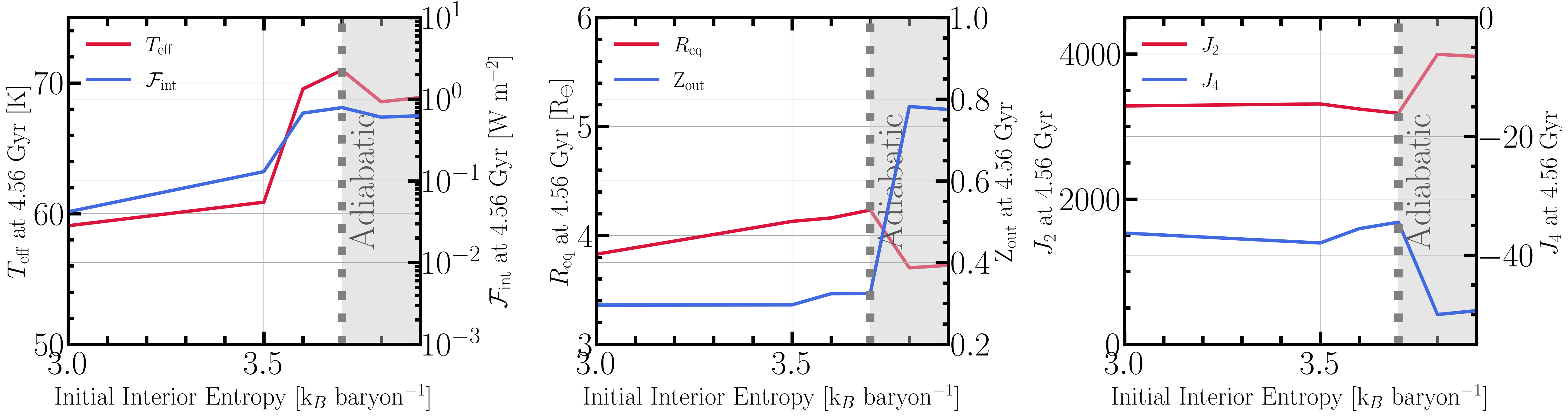}
\caption{Present-age parameters as a function of the initial interior entropy for the models shown in Figure \ref{fig:fig3}. The effective temperature and luminosities increase with higher initial interior entropy until a threshold of $\sim$3.7 k$_B$ baryon$^{-1}$ renders the evolution adiabatic. The threshold values depend on the ice mixtures used (4:1:7 methane-ammonia-water chosen here), but it nevertheless occurs with different ice mixtures and even pure water (see discussion at the end of Section \ref{sec:experiments}). Once ``hotter'' models become adiabatic, they accumulate more heavy metals (center panel, blue line) and display smaller radii (center panel, red curve). Note that adiabatic evolution post-mixing can have a similar luminosity output to unmixed models (left panel), but show significantly different radii and gravitational harmonics (right-most panel).}
\label{fig:fig9}
\end{figure*}

Figure \ref{fig:fig9} shows an example of the correlation of fraction errors of the gravity harmonic values for models in the parameter search described in Section \ref{sec:evolution}. These present-day values from the evolution models are compared with the observations, showing that greater amounts of rocks are needed to satisfy both $J_2$ and $J_4$ simultaneously. Since Uranus models are hypothesized here to maintain their primordial heat, their radii remain high without the presence of rocks in their envelopes. We acknowledge that a more formal and systematic statistical study is warranted, but we leave this as a topic for future work.

\begin{figure*}[ht!]
\centering
\includegraphics[width=\textwidth]{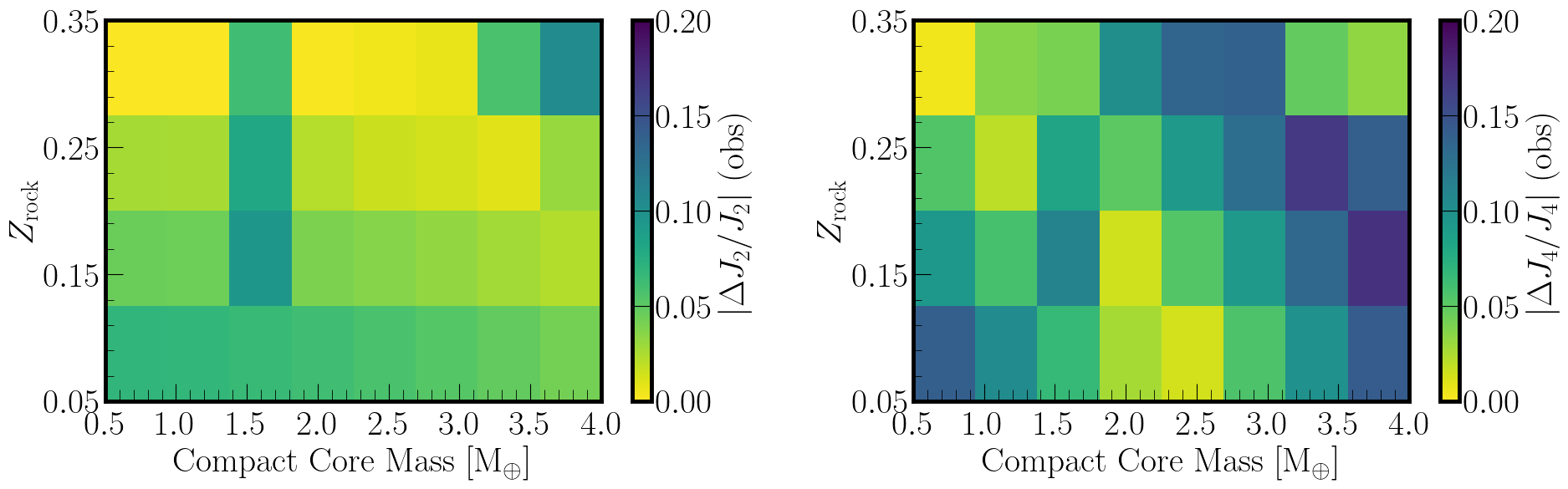}
\caption{Example fractional error mapping of present-age $J_2$ (left panel) and $J_4$ (right panel) of Uranus-mass models with respect to the amount of rocks in the envelope. These models are part of the parameter search described in Section \ref{sec:evolution}. For simplicity, and for this example, the outer envelope mass fraction is held constant at 0.35, and the moment of inertia factor is held constant at 0.2222. Higher rock envelope mass fractions ($Z_{\rm rock}$) are systematically preferred by $J_2$, though a combination of compact core masses with $Z_{\rm rock}$ are possible to satisfy $J_4$. When accounting for both $J_2$ and $J_4$ together, the upper right corner yields the least errors.}
\label{fig:fig10}
\end{figure*}

\clearpage

\bibliography{references}{}
\bibliographystyle{aasjournal}

\end{document}